\documentclass[12pt,twoside]{article}
\usepackage{latexsym}
\usepackage{longtable}
\usepackage{epsfig}
\usepackage{graphicx,psfrag}
\usepackage{amssymb,amsmath,amsthm,booktabs,mathtools}
\usepackage{bm}
\usepackage{color}
\usepackage{subfig}
\usepackage{dutchcal}

\numberwithin{equation}{section}

\newtheorem{rema}{Remark}[section]

\newcommand{\al}[1]{\begin{align} 
#1
\end{align} }

\newcommand{\rndb}[1]{  \left( #1 \right)  }

\newcommand{\bc}{\begin{center}}
\newcommand{\ec}{\end{center}}
\def\ba#1{\begin{array}{#1}\displaystyle}
\newcommand{\ea}{\end{array}}

\newcommand{\beq}{\begin{equation}}
\newcommand{\eeq}{\end{equation}}
\newcommand{\beqa}{\begin{eqnarray}}
\newcommand{\eeqa}{\end{eqnarray}}

\newcommand{\n}{\nonumber\\}
\newcommand{\bi}{\begin{itemize}}
\newcommand{\ei}{\end{itemize}}

\def\lt#1{\left#1}
\def\rt#1{\right#1}
\def\t#1{\tilde{#1}}
\def\h#1{\hat{#1}}
\def\b#1{\bar{#1}}
\def\frc#1#2{\frac{#1}{#2}}

\newcommand{\p}{\partial}

\newcommand{\bra}{\langle}
\newcommand{\ket}{\rangle}
\newcommand{\Z}{{\mathbb{Z}}}

\newcommand{\R}{{\mathbb{R}}}

\newcommand{\Or}{{\cal O}}

\newcommand{\ep}{\epsilon}

\newcommand{\ri}{{\rm i}}
\newcommand{\re}{{\rm e}}
\newcommand{\rl}{{\rm l}}
\newcommand{\rr}{{\rm r}}
\newcommand{\dd}{{\rm d}}
\DeclareMathOperator{\sgn}{sgn}
\newcommand{\rre}{{\rm rest}}
\def\losymbol#1{\mathcal{#1}}

\newcommand{\halmos}{\rule{1ex}{1.4ex}}
\newcommand{\eproof}{\hspace*{\fill}\mbox{$\halmos$}}

\begin{document}

\begin{center}
{\Large {\bf Fluctuations in ballistic transport\\[0.2cm] from Euler  hydrodynamics}}

\vspace{1cm}

{\large Benjamin Doyon and Jason Myers}
\vspace{0.2cm}

{\small\em
Department of Mathematics, King's College London, Strand, London WC2R 2LS, U.K.}
\end{center}
\vspace{1cm}

\noindent We propose a general formalism, within large deviation theory, giving access to the exact statistics of fluctuations of ballistically transported conserved quantities in homogeneous, stationary states. The formalism is expected to apply to any system with an Euler hydrodynamic description, classical or quantum, integrable or not, in or out of equilibrium. We express the exact scaled cumulant generating function (or full counting statistics) for any (quasi-)local conserved quantity in terms of the flux Jacobian. We show that the ``extended fluctuation relations" of Bernard and Doyon follow from the linearity of the hydrodynamic equations, forming a marker of ``freeness" much like the absence of hydrodynamic diffusion does. We show how an extension of the formalism gives exact exponential behaviours of spatio-temporal two-point functions of twist fields, with applications to order-parameter dynamical correlations in arbitrary homogeneous, stationary state. We explain in what situations the large deviation principle at the basis of the results fail, and discuss how this connects with nonlinear fluctuating hydrodynamics. Applying the formalism to conformal hydrodynamics, we evaluate the exact cumulants of energy transport in quantum critical systems of arbitrary dimension at low but nonzero temperatures, observing a phase transition for Lorentz boosts at the sound velocity.
\vspace{1cm}

{\ }\hfill
\today

\tableofcontents

\section{Introduction}
Far-from-equilibrium physics has seen a large amount of theoretical and experimental developments in recent years \cite{trove.nla.gov.au/work/7037197,kinoshita,Jezouin601,Brantut713,eisertreview}.  A distinctive feature of such nonequilibrium states is entropy production and the breaking of time-reversal invariance. This is associated with the existence non-zero currents describing transport of various quantities, such as particles, charge or energy. These effects are ubiquitous in nature and of fundamental importance. Despite this, there exists no fully satisfying non-equilibrium parallel to equilibrium thermodynamics, that is, organising principles for the behaviour and statistics of non-equilibrium currents. A promising avenue is the study of fluctuations, and a widely popular approach is the use of large deviation theory (LDT) \cite{TOUCHETTE20091,inbookTH}, which studies the rare but significant fluctuations around almost-sure values of macroscopic quantities. In particular, fluctuations in transport encode many universal properties of non-equilibrium physics, see \cite{MARCONI2008111,RevModPhys.81.1665,1742-5468-2007-07-P07023,1742-5468-2011-01-P01030,1751-8121-48-50-503001}. LDT offers a conceptual link with thermodynamics, and gives a general understanding of a wide class of nonequilibrium phenomena.

In the context of nonequilibrium transport, it is natural to focus on the LDT for the total transfer of quantities, at long times, between two or more macroscopic regions. As an illustration consider the total energy that has passed from the left to the right halves of an infinitely large system after a long time $t$; one can enquire about the distribution of this random variable as $t\to\infty$. Of particular interest are the cumulants, scaled by $1/t$, of the transferred quantity in the large-$t$ limit. These are finite if the expected large deviation principle holds, and are encoded within the scaled cumulant generating function (SCGF), or full counting statistics, a non-equilibrium counterpart to the equilibrium free energy. The LDT for transport has been studied in many systems, with the SCGF often calculated exactly. For quantum transport of free fermions, the SCGF of $U(1)$ charges is given by the celebrated Levitov-Lesovik formula \cite{levles,Avron2008}, which has applications in mesoscopic physics. Free-particle advanced techniques have been used \cite{FullCountingStatisticsResonantLevelModel,PhysRevLett.99.180601,DLSB,GawTaubNonEq2015,Y18,Moriya2019}, see also \cite{Bernard-Doyon-timeReversal}, and exact results exist in certain integrable impurity models \cite{PhysRevLett.107.100601} and in general 1+1-dimensional conformal field theory (CFT)  \cite{1751-8121-45-36-362001,Bernard2015}, see the review \cite{1742-5468-2016-6-064005}. These systems admit ballistic transport, and nonequilibrium currents are generated by the partitioning protocol \cite{Spohn1977,Ruelle2000,JaksicPilletMathTheor} (see also \cite{1742-5468-2016-6-064005} and references therein), where an imbalance exists in the initial condition. There are also many exact results for the SCGF in classical stochastic many-body systems such as ``exclusion processes", see e.g.~\cite{1742-5468-2007-07-P07023,1742-5468-2011-01-P01030,1751-8121-48-50-503001}. Many techniques have been used, and a successful framework is macroscopic fluctuation theory (MFT) \cite{PhysRevLett.87.150601,PhysRevLett.87.040601,bertinietal,PhysRevLett.92.180601,RevModPhys.87.593}, based on a hydrodynamic description and taking as input diffusion coefficients. Exact results for fluctuations in open or stochastic quantum systems have also been obtained, see e.g. \cite{PhysRevLett.112.067201,Znidaric2014a,Znidaric2014b,PhysRevE.96.052118,BBJ17,BBJ18}, and, recently, in certain cellular automata \cite{BucaGarrahanProsenVanicat19}.

In this paper, we propose a general theory for studying the LDT of {\em ballistic transport} in {\em homogeneous, stationary, maximal entropy states} (MES) of deterministic many-body systems. MES are temporally and spatially homogeneous states (thermodynamic ensembles) reached after relaxation processes have occurred \cite{eisertreview}. In such states, entropy is maximised with respect to all available local conservation laws. They include thermal Gibbs states, Galilean or relativistic boosts thereof, and, in integrable systems, generalised Gibbs ensembles \cite{PhysRevLett.98.050405,PhysRevLett.115.157201,Doyon2017,1742-5468-2016-6-064002} (experimentally observed \cite{Langen207}). MES may admit ballistic currents if there are conserved charges which are odd under time reversal, such as the momentum; in such cases, these are nonequilibrium steady states. In particular, they include the nonequilibrium steady states emerging in the partitioning protocol.

In the setup we consider, time evolution is deterministic, and initial states are fluctuating. In a hydrodynamic description \cite{Spohn-book}, the presence of ballistic transport leads to nontrivial Euler-scale hydrodynamic equations, and to Euler-scale linear fluctuating hydrodynamics. We use aspects of these, see \cite[App A]{SpohnNonlinear}, \cite{1742-5468-2015-3-P03007} and \cite{SciPostPhys.3.6.039}, in order to propose a framework for the LDT of ballistic transport, expressing exact SCGFs in terms of the flux Jacobian (or linearised Euler matrix). Our proposal somewhat parallels MFT, being based on a hydrodynamic description, but differs from MFT in that it only necessitates the Euler scale -- diffusion and other higher-order contributions would give subleading corrections. It can be applied to a large variety of systems with Euler hydrodynamics, quantum or classical, integrable or not, interacting or not.

The theory is based on biasing the measure by a total time-integrated current, thus accessing rare fluctuations and explicitly generating the scaled cumulants. Such a bias is a widely used technique in stochastic systems (sometimes referred to as exponential tilting, or $s$-ensemble), and how it gives rise to a new stochastic dynamics is referred to as the (classical or quantum) generalised Doob transform, see \cite{Chetrite2015,CGLPE18} and references therein. Here, instead, we find the exact modification of the initial state that reproduces this bias.

The proposal provides an organising principle for all results for exact SCGF in homogeneous free particle models and 1+1-dimensional CFT; it is a nonlinear generalisation of principles found for these systems \cite{Bernard-Doyon-timeReversal,1751-8121-45-36-362001,Bernard2015,1742-5468-2016-6-064005}. We clarify the origin of the ``extended fluctuation relations'' of Bernard and Doyon \cite{Bernard-Doyon-timeReversal}, showing that they arise when the flux Jacobian, in the coordinates of the conserved densities, is state-independent (the hydrodynamic equations are linear). We propose that this is a property of a many-body system that characterises it as being ``free" by opposition to ``interacting", much like the absence of diffusion is \cite{SpohnInteracting}.

The theory generalises to the total integrated currents along arbitrary rays (space-time points $\{(x,t) : x/t=\xi\}$). We show how this gives the exact exponential asymptotics of spatio-temporal two-point correlation functions of twist fields, and thus of order-parameter dynamical correlation lengths in homogeneous, stationary states.

In certain situations, we find divergent scaled cumulants, signalling that the fluctuations scale in a different fashion. Seeing the SCGF as a function of the state's parameters (or of the bias), this may be interpreted as a ``dynamical phase transition," of the type seen in other contexts, see e.g.~\cite{Bodineau2005,Garragan2007,Garragan2009,Espigares2013,Jack2014,Hurtado2014,Tsobgni2016,Tizon2017,Lazarescu2017}. We explain why this occurs from hydrodynamic principles, and how this may connect with the breaking of Gaussianity found in nonlinear fluctuating hydrodynamics \cite{SpohnNonlinear,1742-5468-2015-3-P03007,ChenDeGierHirikiSasamotoFluctuHydro18}.

Of immediate importance is the application to integrable systems, a detailed examination of which is completed in a separate work \cite{Myers-Bhaseen-Harris-Doyon}. There, we exploit generalised hydrodynamics \cite{PhysRevX.6.041065,PhysRevLett.117.207201,DY,dNBD,dNBD2} (experimentally verified \cite{Schemmerghd}), and confirm the proposal in the hard rod gas by comparing with Monte Carlo simulations. In the present paper, we apply the theory to non-integrable quantum critical systems of dimension higher than 1, using conformal hydrodynamics. Nonequilibrium steady states for energy transport were first studied in such systems in \cite{BDLS,CKY} and \cite{Pour,LSDB,SH}. Here we obtain exact fluctuation results. We write explicitly the first few scaled cumulants for energy transport as functions of the rest-frame temperature and the relativistic boost, and differential equations for the SCGF, which we solve numerically. We observe a dynamical phase transition in thermal states boosted to the sound velocity.

The paper is organised as follows. In section \ref{sectcontext}, we explain the context and review the main aspects of large deviation theory for transport. In section \ref{sectmain}, we present our main results, explain the main idea of the derivation, discuss the extended fluctuation relations, give the generalisation to arbitrary rays and the application to twist-field correlation functions, and discuss dynamical phase transitions. In section \ref{sectCFT} we present the application to conformal hydrodynamics, and in section \ref{sectconclu} we present conclusions and open questions. Finally, in Appendix \ref{appder} we provide the main derivation of the general results, in Appendix \ref{appeuler} we review basic aspects of Euler hydrodynamics, including the solution to the Riemann problem in free (linear) hydrodynamics and the normal modes of conformal hydrodynamics, and in Appendix \ref{appmany} we discuss the multi-parameter SCGF and present related general arguments.

\section{Maximal entropy states and large deviation theory}

\label{sectcontext}

In this section, we first describe the context in which the main results are proposed to apply: all systems that possess an Euler hydrodynamic description. We specify what we believe would be the general properties of many-body systems that are expected to be necessary for our results to hold. We then recall the main aspects of large deviation theory that we need, including the scale cumulant generating function (SCGF).

\subsection{Systems and states of interest}
\label{ssectmes}

We consider an infinite-length one-dimensional many-body system with a dynamics that is homogeneous both in space and time, and with local interactions. This can be a classical or quantum lattice model, field theory or gas, and the dynamics may or may not be generated by a Hamiltonian. We believe the results apply both to deterministic and stochastic dynamics, although we concentrate on the former\footnote{In the stochastic case, the dynamics should be Markovian, and the various concepts used here have natural stochastic correspondents, for instance conserved quantities should be interpreted as martingales.}, where randomness lies in the initial state. The assumption of one-dimensionality may be partially lifted by applying the results below to effectively one-dimensional transport in higher-dimensional systems; we discuss this in the context of higher-dimensional conformal hydrodynamics in section \ref{sectCFT}.

The model is assumed to admit a certain number of homogeneous conserved charges $Q_i = \int_\R \dd x\,\losymbol{q}_i(x,t)$. These are dynamical observables satisfying $\dd Q_i/\dd t=0$. They are assumed to have associated conservation laws
\beq\label{conslaw}
	\p_t \losymbol{q}_i(x,t) + \p_x \losymbol{j}_i(x,t) = 0
\eeq
indexed by $i$. Here $\losymbol{q}_i(x,t)$ and $\losymbol{j}_i(x,t)$ are the charge density and current, respectively, at space-time point $(x,t)$, which are local (supported on a finite region containing $x$) or quasi-local (an appropriate extension \cite{IlievskietalQuasilocal,Doyon2017}) observables at that point (we likewise say that $Q_i$ are local or quasi-local charges). The conservation laws follow purely from the dynamics of the model. If the system is Hamiltonian, then the conserved charges include the Hamiltonian, and we assume them to be in involution (they commute with each other as well as with the Hamiltonian).

States are statistical in nature; fluctuations in deterministic dynamical systems may come from various sources, such as fluctuating initial conditions or quantum fluctuations. Following the general philosophy of the $C^*$-algebra description of quantum and classical statistical mechanics, states are fully described by the set of all expectations, here denoted using the bracket notation $\bra\cdots\ket$, of local observables; this space may be completed under various norms, see e.g.~\cite{IsraelConvexity,BratelliRobinson12,Doyon2017}.

We consider the manifold of homogeneous and stationary (i.e.~invariant under space and time translations) {\em maximal entropy states} (MES). Formally, MES are characterised by as many Lagrange parameters $\beta^i$ (indexed by $i$) as there are conserved quantities $Q_i$, and have probability measure or density matrix proportional to $\re^{-\sum_i \beta^i Q_i}$. We will denote expectations in a MES by $\bra\cdots\ket_{\underline\beta}$, where $\underline\beta$ is the vector of all $\beta^i$'s, coordinates for the MES manifold (when it is clearer, we will also use the notation $\beta^\bullet$ to represent the set of $\beta^i$'s for all $i$).

The form $\re^{-\sum_i \beta^i Q_i}$ for the probability measure or density matrix is formal in most situations. When the series $\sum_i \beta^i Q_i$ truncates and the charges $Q_i$ are local, there is a variety of ways to make it rigorous in the context of $C^*$ algebras: as an infinite-volume limit; via the Kubo-Martin-Schwinger (KMS) relation (in the quantum case), or the Dobrushin-Lanford-Ruelle (DLR) equations (in the classical case); by a precise notion of entropy maximisation; or by considering appropriate tangents to a manifold of states; see \cite{IsraelConvexity,BratelliRobinson12} for discussions. A formulation related to the latter, which is developed in \cite{Doyon2017}, accounts for ``generalised thermalisation" in integrable systems, where the series $\sum_i \beta^i Q_i$ may be infinite and the charges quasi-local. It is based on considering conserved charges $Q_i$ as vectors in the tangent space to the MES manifold. A charge $Q_i$ deforms the state according to the equation
\beq\label{dbetaeq}
	-\frc{\p}{\p\beta^i} \bra \Or(0,0) \ket_{\underline\beta} = (\losymbol{q}_i,\Or)_{\underline\beta}
\eeq
where the symmetric inner product on the space of local fields is
\beq\label{innerproduct}
	(\Or,\Or')_{\underline\beta} = \int_{-\infty}^\infty \dd x\,\bra \Or(x,0),\Or(0,0)\ket_{\underline\beta}^{\rm c}
\eeq
and
\beq\label{connected}
	\bra \Or(x,t),\Or'(0,0)\ket^{\rm c}_{\underline\beta} =
	\frc12 \big\bra \Or(x,t)\Or'(0,0) +
		\Or'(0,0)\Or(x,t) \big\ket_{\underline\beta} 
	- \bra \Or(x,t)\ket_{\underline\beta}
	\bra \Or'(0,0)\ket_{\underline\beta}
\eeq
refers to the connected, symmetrised correlation function\footnote{Here and below, we assume all observables to be real (hermitian) [in the classical (quantum) case]. The symmetrisation used in the first term on the right-hand side of \eqref{connected} is a natural prescription in order to take care of the lack of commutativity in the quantum case, see for instance  \cite{Doyon2017}.}. According to \cite{Doyon2017}, the tangent space is the Hilbert space, induced by \eqref{innerproduct}, of pseudolocal conserved charges \cite{ProsenPseudo1,ProsenPseudo2,IlievskietalQuasilocal}, and we believe that the results below hold if the set $Q_i$ is complete in the sense that it spans a dense subset of this Hilbert space. In one dimension, away from the ground state, there is no spontaneous symmetry breaking, and thus MES describe ``pure phases". As such, they are extremal states \cite{IsraelConvexity,BratelliRobinson12}, and therefore cluster at large distances. Clustering of two-point functions can be shown to be exponential on general grounds in Gibbs states of local quantum hamiltonians \cite{Araki}. In any case, below we assume clustering to be strong enough so that integrals of connected correlation functions converge.

The manifold of MES forms the basis of the emergent hydrodynamics in slowly varying, long-wavelength states \cite{Spohn-book}, in that it gives rise to the ``equations of state": the relation between average currents and average densities. 

In generic Galilean invariant quantum and classical gases, the set of conserved quantities contain the particle number, the energy and the momentum, and the MES are simply the Gibbs ensembles and Galilean boosts thereof. Similar statements hold for generic relativistic gases. In these cases, the Euler hydrodynamics is the standard one, Galilean or relativistic. In integrable systems, there are infinitely many conserved quantities, and the MES manifold is infinite dimensional. In these cases, the MES are referred to as generalised Gibbs ensembles \cite{PhysRevLett.98.050405,PhysRevLett.115.157201,Doyon2017,1742-5468-2016-6-064002}, and the Euler hydrodynamics, referred to as generalised hydrodynamics, was developed in \cite{PhysRevX.6.041065,PhysRevLett.117.207201,DY} (and was recently verified experimentally \cite{Schemmerghd}). In a large family of integrable models, the Lagrange parameters $\beta^i$'s are more appropriately represented by a function on a ``spectral space", a space of available stable ``quasi-particles", via the thermodynamic Bethe ansatz \cite{Yang-Yang-1969,ZAMOLODCHIKOV1990695,TakahashiTBAbook}, see e.g. \cite{IlievskietalQuasilocal,IlievskiInteracPart}; thus in these cases the MES manifold is a manifold of functions, whose full description is however not yet known in most models.

\begin{rema} MES are essentially the ``invariant equilibrium states" as first developed by Gibbs. Gibbs' theory for gases usually does not include the momentum and its associated intensive ``potential" as thermodynamic quantities, but these may be simply re-introduced by Galilean (or relativistic) boosts. Away from the rest-frame, the state is no longer time-reversal invariant, hence the appellation ``equilibrium state" is not appropriate, and ``maximal-entropy states" seems better suited to describe the full set. In more general systems, such as in integrable systems, there are usually infinitely-many conserved charges that break time-reversal invariance, which cannot be accounted for simply by Galilean or relativistic boosts.
\end{rema}

\begin{rema}\label{remamore}
Physically, the MES manifold is the set of states that are homogeneous, stationary and clustering, and that describe averages of observables on finite regions emerging after relaxation. That is, these are all states that occur after evolving for a long time from generic initial states in infinite volume, in accordance with local relaxation in isolated, thermodynamically large systems \cite{eisertreview}. The set of conservation laws restricts the state's evolution, and MES are thus states where entropy is maximised with respect to all available local (and quasi-local) conservation laws. An interesting question, which to our knowledge has not been settled, is as to if the set of homogeneous, stationary and clustering states includes other states than MES.
\end{rema}

\subsection{Large deviations in transport}

In LDT one concentrates on fluctuating quantities $J^{(t)}$, which are extensive with respect to some parameter $t$, and whose densities $J^{(t)}/t$ take almost-sure values $\b\j$ in the limit $t\to\infty$. For the purpose of studying nonequilibrium transport, it is natural to focus on the LDT of the {\em total transfer} of a conserved quantity after time $t$, between two regions of a system. In higher-dimensional systems, we assume transport to occur in a single direction of space, so that the system is effectively one-dimensional. Transfer occurs say from the left, $x<0$, to the right, $x>0$. The conserved quantity $Q$ is one of the $Q_i$'s, with index $i=i_*$,
\beq\label{istar}
	Q = Q_{i_*},\qquad \losymbol{q}(x,t) = \losymbol{q}_{i_*}(x,t),
	\qquad \losymbol{j}(x,t) = \losymbol{j}_{i_*}(x,t)	.
\eeq
The total transfer of $Q$ after time $t$, for the purpose of the LDT, is the total current passing by the origin,
\beq\label{Qt}
	J^{(t)} = \int_0^t \dd s\,\losymbol{j}(0,s).
\eeq

The quantity $J^{(t)}$ is a random variable, whose probability distribution $\mathbf{P}$ is determined by the state of interest, which we take to be a MES for some $\underline\beta$. According to the large deviation principle, such extensive quantities have probability distributions that are exponentially peaked at the almost-sure value (here the scaling in time has exponent 1, which is the one relevant here)\footnote{\label{fn} Following standard notation, $A(t)\asymp B(t)$ means $\lim_{t\to\infty} (\log A(t))/(\log B(t))=1$.},
\beq\label{ld}
	\mathbf{P}(J^{(t)}=tj) \asymp e^{-t I(j)},\qquad I(\b\j) = 0,\qquad I(j)>0\quad(j\neq \b\j).
\eeq
The almost-sure value $\b\j$ is simply the average current in the state of interest. The function $I(j)$ controlling this exponential is referred to as the large-deviation function. It describes the probabilities of rare but significant events where the quantity $J^{(t)}$ deviates by large amounts from $t\b\j$. Its Legendre-Frenchel transform, $F(\lambda)$, is the scaled cumulant generating function (SCGF) for $J^{(t)}$,
\beq\label{scgf}
	F(\lambda) = \lim_{t\to\infty} t^{-1}\log \bra \re^{\lambda J^{(t)}}\ket_{\underline\beta} =
	 \sum_{n=1}^{\infty} \frc{\lambda^n}{n!} c_n.
\eeq
Here $c_n$ are the cumulants, scaled by time. Up to a conventional minus sign, the function $F(\lambda)$ can be interpreted as a nonequilibrium equivalent of the equilibrium specific free energy, with nonequilibrium partition function $Z = \bra \re^{\lambda J^{(t)}}\ket_{\underline\beta} \asymp e^{tF(\lambda)}$. Crucially, the large-deviation principle \eqref{ld} implies that all cumulants of the random variable $J^{(t)}$ scale like $t$ at large $t$. An important question in LDT is the evaluation of the exact large deviation function $I(j)$, or the exact SCGF $F(\lambda)$.

\begin{rema} In the LDT for quantum systems, formulating the problem of the fluctuations of the total transfer of a charge $Q$ by a direct interpretation of the above formulae is not physically natural: the operator $\int_0^t \dd s\,\losymbol{j}(0,s)$ is not a natural quantum observable on which von Neumann measurements can be made, as it involves a time integral. Instead, one needs a formulation that takes into account properly the quantum nature of the system, see e.g.~\cite{RevModPhys.81.1665}. This can be via a two-time von Neumann measurement procedure where the charge difference between the left and right halves of the system is measured at time 0, the system is let to evolve and the charge difference is again measured at time $t$. It can also be via some indirect measurement scheme, for instance where the current passing by the origin is coupled to an external device on which von Neumann measurements are made, see e.g. \cite{levles}. There are indications that suggest that different measurement schemes lead, in the large deviation limit and for ballistic transport, to the same result, and that this result is in agreement with \eqref{Qt}, \eqref{ld}, and \eqref{scgf}, see e.g. the review \cite{1742-5468-2016-6-064005}. In this paper, we assume this to be the case.
\end{rema}

\section{Main results: ballistic large deviation theory}

\label{sectmain}

In this section, we give the exact expressions for SCGF for ballistic transport in terms of objects from linear fluctuating hydrodynamics. We present an overview of the derivation, with the details given in Appendix \ref{appder}. We then provide various consequences, and we extend the results to fluctuations along arbitrary rays, with applications to correlation functions of twist fields. 

\subsection{Scaled cumulant generating function for ballistic transport}\label{ssectmain}

It is expected that there is a set of averages of all local or quasi-local densities, which we will denote by $\mathsf{q}_i = \bra \losymbol{q}_i(0,0)\ket_{\underline\beta}$, which provide a good system of coordinates for the MES manifold -- that is, the map $\underline\beta \mapsto \underline{\mathsf{q}}$ is bijective (from an appropriate space of $\underline\beta$). In (a large family of) integrable systems, this set is, again, more appropriately represented by a function on a spectral space (the ``quasi-particle density" of the thermodynamic Bethe ansatz \cite{Yang-Yang-1969,ZAMOLODCHIKOV1990695,TakahashiTBAbook}).

Consider, then, the averages of the currents, $\mathsf{j}_i = \bra \losymbol{j}_i(0,0)\ket_{\underline{\beta}}$, as functions of the state coordinates $\underline{\mathsf{q}}$. These model-dependent functions -- the fluxes -- are the equations of state of the model. Construct the flux Jacobian (or linearised Euler matrix)
\beq\label{Amatrix}
	\mathsf A_{i}^{~j} = \frc{\p\mathsf{j}_i}{\p{\mathsf{q}_j}}.
\eeq
This is a (model-specific) matrix that is a function of the MES, hence a function of the coordinates $\underline{\beta}$, or $\underline{\mathsf{q}}$. The flux Jacobian is at the basis of the Euler hydrodynamic theory (see Appendix \ref{appeuler}), and is a fundamental part of what is often referred to as linear fluctuating hydrodynamics.

Consider the MES characterised by some coordinates $\underline\beta$. We define a flow $\lambda\mapsto \underline\beta(\lambda)$ on the manifold of MES, starting on this state $\underline\beta(0)=\underline\beta$ and for $\lambda$  lying in some interval of $\R$, by the differential equation
\beq\label{flowbeta}
	\frc{\dd}{\dd\lambda}\mathsf{\beta}^i(\lambda) = -\sgn(\mathsf A(\lambda))_{i_*}^{~i},
\eeq
where $\mathsf A(\lambda)$ is the flux Jacobian in the state with Lagrange parameters $\underline\beta(\lambda)$. On the right-hand side, the sign of the flux Jacobian $\mathsf A$ is the matrix obtained by diagonalising $\mathsf A$ and taking the sign of its eigenvalues,
\beq\label{diag}
	\sgn(\mathsf A) = \mathsf M\sgn(v^{\rm eff})\mathsf  M^{-1},\qquad \mathsf A = \mathsf Mv^{\rm eff} \mathsf M^{-1}
\eeq
where $v^{\rm eff}={\rm diag}(v^{\rm eff}_1,v^{\rm eff}_2,\ldots)$ is the diagonal matrix of eigenvalues (the $v^{\rm eff}_i$'s are the effective velocities attached to the normal modes of the hydrodynamics, see \eqref{veffdiagq}). Recall that $i_*$ is the index corresponding to the conserved charge of interest, Eq.~\eqref{istar}.

By the chain rule, Eq.~\eqref{flowbeta} is equivalent to the following flow in the conserved density coordinates,
\beq\label{flow}
	\p_{\lambda}\mathsf{q}_i(\lambda) = \big(\sgn(\mathsf A(\lambda))\,\mathsf C(\lambda)\big)_{i_*i}.
\eeq
The static correlation matrix is defined by
\beq\label{Cmatrix}
	\mathsf C_{ij} = (\losymbol{q}_i,\losymbol{q}_j)_{\underline\beta} = -\frc{\p\mathsf{q}_j}{\p{\beta^i}},
\eeq
and $\mathsf C(\lambda)$ in \eqref{flow} is the flow-dependent static correlation matrix, evaluated in the state with Lagrange parameters $\underline\beta(\lambda)$.

Our main result, shown in Appendix \ref{appder}, is as follows. We identify the flow parameter $\lambda$ with the conjugate parameter in \eqref{scgf}, and we have an expression for the SCGF $F(\lambda)$ directly in terms of the current along the flow:
\beq\label{scgfres}
	F(\lambda) = \int_0^\lambda \dd\lambda'\,\mathsf{j}(\lambda')
\eeq
where $\mathsf{j}(\lambda) =  \mathsf{j}_{i_*}(\lambda) = \bra \losymbol{j}_{i_*}(0,0)\ket_{\underline{\beta}(\lambda)}$. Thus, the knowledge of the Euler hydrodynamics (giving the flux Jacobian $\mathsf A$, and the currents $\mathsf{j}_i$, as functions of the state) is sufficient in order to obtain the SCGF. As we also show in Appendix \ref{sappc2}, this agrees with, and largely generalises, the result for the second cumulant $c_2$ which follows from the current-current sum rule written in \cite{1742-5468-2015-3-P03007,SciPostPhys.3.6.039}.

If $F(\lambda)$ is strictly convex and everywhere differentiable, then the Legendre-Frenchel transform reduces to the Legendre transform, and it is a simple matter to obtain the large-deviation function as:
\beq
	I(j) =j \lambda(j)
	- F(\lambda(j)),\qquad \mathsf{j}(\lambda(j)) = j.
\eeq

\begin{rema} \label{remaext} One can, formally, generalise to SCGFs with multiple parameters $\lambda^j$ associated to all currents $\losymbol{j}_j$, with
\beq
	F(\lambda_1,\lambda_2,\ldots) = \lim_{t\to\infty} t^{-1}\log \Big\bra \exp{\sum_j \lambda_j \int_0^t \dd s\,\losymbol{j}_j(0,s)}
	\Big\ket_{\underline\beta}.
\eeq
In \eqref{flow} and \eqref{flowbeta} we make the replacements $\lambda\mapsto\lambda^j$ and $i_*\mapsto j$. This however requires multiple differentiability with respect to the parameters $\lambda_j$, which are nontrivial relations on the matrix $\mathsf A$ as a function of these parameters. In this paper we do not investigate this aspect, however see Appendix \ref{appmany}.
\end{rema}

\begin{rema}
In certain cases, where the generating function of the currents separates into a sum of functions of the normal modes, we opbtain a more explicit expression for $F(\lambda)$, developed in Appendix \ref{appmany}. This agrees with the general expression found in integrable systems, see \cite{Myers-Bhaseen-Harris-Doyon}.
\end{rema}

\subsection{Derivation: biasing the measure}\label{ssectskewing}

The derivation of the main results \eqref{scgfres} with \eqref{flowbeta} is provided in Appendix \ref{appder}. The main argument is to bias the measure in a particular way, and show that the bias generates a flow on the MES manifold. This latter fact can be shown either using the theory of pseudolocal charges \cite{IlievskietalQuasilocal,Doyon2017}, or from a strong version of the hydrodynamic projection principles \cite{Spohn-book,SpohnNonlinear,1742-5468-2015-3-P03007,SciPostPhys.3.6.039}. In order to specify the explicit flow in terms of the flux Jacobian, one needs certain basic results from linear fluctuating hydrodynamics. Here we give the main lines of the proof. We note that the basic techniques leading to the form \eqref{scgfres} for the SCGF were introduced in \cite{1751-8121-45-36-362001,Bernard2015} in the context of 1+1-dimensional conformal field theory and more generally in \cite{Bernard-Doyon-timeReversal}; in fact, the result \eqref{scgfres} with \eqref{flowbeta} may be seen as a nonlinear generalisation of the results found there, as is made clearer in the next subsection.

Besides assumptions which are expected to be valid quite generally in local many-body systems, the leading assumption of physical relevance is that of {\em sufficiently strong clustering of local observables at long times}. That is, multi-point connected correlation functions of local fields, in particular of local currents, vanish at large time separations, in a way that makes them integrable. Although this vanishing is expected generically, contrary to that of connected correlation functions at large space separations, it is nevertheless not guaranteed in MES. Its breaking leads to a failure of the large deviation principle \eqref{ld}. We discuss the physics and potential consequences of this in subsection \ref{ssectbreaking}.

Let us modify the measure for the state $\bra\cdots\ket_{\underline\beta}$ by a time-integral of the current $\losymbol{j}(0,t)$. That is, let us construct a family of states $\bra \cdots\ket^{(\lambda)}$, parametrised by $\lambda\in\R$, with $\bra\cdots\ket^{(0)}=\bra\cdots\ket_{\underline\beta}$, obtained by modifying the state $\bra\cdots\ket_{\underline\beta}$ by the insertion of the time-integrated local current $\losymbol{j}(0,t)$ of the charge $Q$ at the origin,
\beq\label{eq0}
	\bra\Or(x,t)\ket^{(\lambda)} = \frc{\bra \re^{\frc{\lambda}2 \int_{-\infty}^\infty \dd t\,\losymbol{j}(0,t)}\Or(x,t)
	\re^{\frc{\lambda}2 \int_{-\infty}^\infty \dd t\,\losymbol{j}(0,t)}
	\ket_{\underline\beta}}{\bra \re^{\lambda \int_{-\infty}^\infty \dd t\,\losymbol{j}(0,t)}\ket_{\underline\beta}}
\eeq
(the symmetrisation guarantees that averages of hermitian observables are real numbers in the quantum case). This is well defined as a formal expansion in $\lambda$ if connected correlation functions vanish fast enough at large time separation, and it is in fact expected to be well defined for real values of $\lambda$ in an interval containing the origin. In particular, we have
\beq\label{derstate}
	\frc{\dd}{\dd\lambda} \bra\Or\ket^{(\lambda)} =
	\int_{-\infty}^\infty \dd t\,\bra \losymbol{j}(0,t),\Or\ket^{(\lambda), \rm c}
\eeq
where the connected correlation function $\bra \cdot,\cdot \ket^{(\lambda), \rm c}$ is defined as in \eqref{connected} but for the state $\bra\cdots\ket^{(\lambda)}$.

As a loose interpretation, the insertion of the exponential of the time integral of the current can be seen as ``biasing" the dynamics, changing the weights of trajectories in order to make rare events ``typical" and access their probabilities. The biasing by a time-integrated current is natural and has been used widely in the study of large deviations in stochastic dynamics or open quantum systems. In this context, one attempts to relate it to a change of the stochastic dynamics or of the Lindbladian, something referred to as the generalised (classical or quantum) Doob transformation \cite{Chetrite2015,CGLPE18}. By contrast, here we relate it to a change of the distribution ruling the initial state, and crucially, the new distribution is still a MES -- we obtain a function $\underline\beta(\lambda)$ with $\underline\beta(0)=\underline\beta$. We determine this change solely from the Euler hydrodynamics of the system.

Before determining $\underline\beta(\lambda)$, we explain how the $\lambda$-dependent state is useful in order to fix $F(\lambda)$; see e.g. \cite{Bernard-Doyon-timeReversal}. Taking the definition \eqref{scgf}, we find, after a convenient shift in the time integration region allowed by stationarity of the state and after symmetrisation,
\beq
	\frc{\dd F(\lambda)}{\dd\lambda} =
	\lim_{t\to\infty} \frc1{t}\int_{-t/2}^{t/2} \dd s\,
	\frc{\bra\re^{\frc{\lambda}2 \int_{-t/2}^{t/2}\dd r\,\losymbol{j}(0,r)}\losymbol{j}(0,s)
	\re^{\frc{\lambda}2 \int_{-t/2}^{t/2}\dd r\,\losymbol{j}(0,r)}\ket_{\underline\beta}
	}{\bra\re^{\lambda \int_{-t/2}^{t/2}\dd r\,\losymbol{j}(0,r)}\ket_{\underline\beta}}.
\eeq
Under an appropriate assumption of sufficiently fast clustering in time, when $t$ is large we expect the main contribution to the $s$ integral to come from the central region away from the boundaries $s=\pm t/2$, where the state is stationary. That is, we may write
\beqa
	\frc{\dd F(\lambda)}{\dd\lambda} &=&
	\lim_{t\to\infty} \frc1{t}\int_{-t/2}^{t/2} \dd s\,
	\frc{\bra\re^{\frc{\lambda}2 \int_{-\infty}^{\infty}\dd r\,\losymbol{j}(0,r)}\losymbol{j}(0,s)
	\re^{\frc{\lambda}2 \int_{-\infty}^{\infty}\dd r\,\losymbol{j}(0,r)}\ket_{\underline\beta}}{\bra\re^{\lambda \int_{-\infty}^{\infty}\dd r\,\losymbol{j}(0,r)}\ket_{\underline\beta}} \n
	&=& \bra \losymbol{j}(0,0)\ket^{(\lambda)}
	\label{Fder}
\eeqa
where we used \eqref{eq0}. The first equality is because the main contribution is from the central region, the second because this contribution is time-independent. Integrating on $\lambda$ with the condition $F(0)=0$, we obtain \eqref{scgfres}.

The state resulting from the $\lambda$-bias is manifestly stationary. Also, by the fact that $\losymbol{j}(x,t)$ is part of a conservation law, we have $\int_{-\infty}^\infty \dd t\,\losymbol{j}(0,t) = \int_{-\infty}^\infty \dd t\,\losymbol{j}(x,t)$ for any $x$, whence the resulting state is homogeneous. It is also possible to argue that the state is clustering. Is this state a MES (see Remark \ref{remamore})? We show in Appendix \ref{appder} that indeed it is. We show in two ways -- either from the theory of pseudolocal charges, or from hydrodynamic projection principles -- that infinitesimal $\lambda$ modifications, eq.~\eqref{derstate}, generate tangents to the MES manifold. Since at $\lambda=0$ the state lies on the MES manifold, then it stays on it.  As a consequence, $\lambda\mapsto \bra\cdots\ket^{(\lambda)}$ forms a path lying within this manifold. Therefore, there exists $\underline\beta(\lambda)$ with $\underline\beta(0)=\underline\beta$ and
\beq\label{princ}
	\bra\Or\ket^{(\lambda)} = \bra\Or\ket_{\underline\beta(\lambda)}.
\eeq
This is the crucial observation of the method.

Using \eqref{derstate}, we then have
\beq\label{eq1}
	\frc{\dd}{\dd\lambda} \bra\Or\ket_{\underline\beta(\lambda)} = \int_{-\infty}^\infty \dd t\,\bra \losymbol{j}(0,t),\Or\ket_{\underline\beta(\lambda)}^{\rm c}
\eeq
for any local or quasi-local observable $\Or$. We may obtain an equation for the coordinates $\mathsf{q}_i(\lambda)$ by specifying $\Or$ in \eqref{eq1} to be the available conserved densities of the model, $\losymbol{q}_i(0,0)$. The left-hand side is therefore the time derivative of the state coordinates. The right-hand side is a time-integrated two-point function of conserved densities and currents, and this is a function of the state, hence can be seen as a function of the state coordinates $\underline{\mathsf{q}}(\lambda)$. Eq.~\eqref{eq1} therefore fully specifies the path by giving its tangent at the point $\underline{\mathsf{q}}(\lambda)$ in terms of a function of $\underline{\mathsf{q}}(\lambda)$. A result from linear fluctuating hydrodynamics \cite{Spohn-book,SpohnNonlinear,1742-5468-2015-3-P03007,SciPostPhys.3.6.039} is that, in an appropriate Euler scaling limit \cite{Spohn-book,SciPostPhys.3.6.039,doyoncorrelations} (see appendix \ref{appder}), where in particular both $x$ and $t$ are large in fixed ratio, we have
\beq\label{jqcorr}
	\bra \losymbol{j}_i(x,t),\losymbol{q}_j(0,0)\ket^{\rm c}_{\underline\beta} \sim (\mathsf A\delta(x-\mathsf At)\mathsf C)_{ij}
\eeq
where $\delta(x-\mathsf At) = \mathsf M\delta(x-v^{\rm eff}t)\mathsf M^{-1}$ (see \eqref{diag}). Integrating over time, the result is independent of $x$, and we obtain 
\beq\label{qjA}
	\int_{-\infty}^\infty \dd t\,\bra \losymbol{j}_i(0,t),\losymbol{q}_j(0,0)\ket^{\rm c}_{\underline\beta} =
	(\sgn(\mathsf A)\,\mathsf C)_{ij}.
\eeq
This is the crucial technical step in the derivation. Equation \eqref{qjA}, and a generalisation of it necessary in order to show that the $\lambda$-bias keeps the state within the MES manifold, is shown more rigorously in Appendix \ref{appder} from hydrodynamic projection. Combining \eqref{qjA} and \eqref{eq1}, we indeed find \eqref{flow}, which implies \eqref{flowbeta}.

\subsection{Constant flux Jacobian and extended fluctuation relations}

The form of the flux Jacobian $\mathsf A_i^{~ j}$ depends on the fluid coordinate system chosen. As the name suggests, the flux Jacobian transforms as a Jacobian: a covariant (contravariant) vector in its first (second) index. Seen as a matrix, this is a similarity transformation, which is generically coordinate-dependent. Therefore, the only coordinate-independent information within the flux Jacobian is its spectrum, the elements of the diagonal matrix $v^{\rm eff}$ in \eqref{veffdiagq} (and in \eqref{diag}).
However, there is more information about the physical system within the flux Jacobian. Indeed, the physical system provides a favoured, special set of coordinate systems: the densities of conserved charges. These are specified by the model up to $\R$-linear transformations, but $\R$-linear transformations form a subset of the set of all coordinate transformations. Hence, one can define a ``natural" flux Jacobian as the flux Jacobian in a system of coordinates given by the conserved densities. This is unique up to $\R$-linear transformations, which are coordinate-independent similarity transformations.

In some cases, for instance in non-interacting models and 1+1-dimensional conformal field theory, the natural flux Jacobian is {\em independent of the state} (a property which is indeed invariant under $\R$-linear transformations). Equivalently, the Euler hydrodynamic equations are linear, see \eqref{eulerquasilinear}.
In many ways, this can be considered as a hydrodynamic system without interactions.

In such cases, it is a simple matter to solve for the flow \eqref{flowbeta}:
\beq\label{solfree}
	\beta^i(\lambda) = \beta^i -\lambda\sgn(\mathsf A)_{i_*}^{~i}.
\eeq
That is, the flow corresponds to a shift of the Lagrange parameters proportional to $\lambda$. In particular, we have from \eqref{scgfres}
\beq\label{scgfefr}
	F(\lambda) = \int_0^\lambda \dd\lambda'\,
	\bra\losymbol{j}\ket_{\{\beta^\bullet -\lambda'\sgn(\mathsf A)_{i_*}^{~\bullet}\}}.
\eeq

Let us consider nonequilibrium steady states emerging from the partitioning protocol \cite{Spohn1977,Ruelle2000,JaksicPilletMathTheor}. In this protocol, two semi-infinite, separate halves (seen as two baths) of the system are initially in different states, often taken to be different thermal states at different temperatures (possibly with different boosts). Suppose the set of Lagrange parameters are $\beta^i_{\rm l}$ (left) and $\beta^i_{\rm r}$ (right). The two halves are then connected to each other and let to evolve for a long time. In any finite region around the connection point, a steady state develops at infinite times, and if ballistic transport is supported, nonequilibrium currents may emerge. Let us denote the Lagrange parameters $\beta^i$ characterising this steady state by $\beta^i(\underline\beta_\rl,\,\underline\beta_\rr)$. In fluid dynamics, this is known as the Riemann problem \cite{BressanNotes}.

In the cases of a natural flux Jacobian that is independent of the state, it is a simple matter to solve the Riemann problem, and to evaluate the steady state in the region around the connection point. We show in Appendix \ref{sappfree} that this solution leads to the relation
\beq\label{efrgeneral}
	\beta^i(\beta_\rl^\bullet,\,\beta_\rr^\bullet)
	-\lambda\sgn(\mathsf A)_{i_*}^{~i}
	=
	\beta^i(\beta_\rl^\bullet - \lambda\delta_{i_*}^\bullet,\,\beta_\rr^\bullet + \lambda\delta_{i_*}^\bullet).
\eeq
This, combined with \eqref{scgfefr}, is the fully general statement of the {\em extended fluctuation relations}, first introduced by Bernard and Doyon \cite{Bernard-Doyon-timeReversal}. That is, according to the extended fluctuation relations, the biasing of the measure necessary to generate transport cumulants can be performed by linear shifts of the Lagrange parameters in the initial baths of the partitioning protocol. Such linear shifts generate fluctuation statistics of the initial state, hence the extended fluctuation relations indicate that, in free models, the statistics of transport fluctuations is directly obtained from that of the initial state fluctuations. This appears to be physically sensible, as without interactions, initial-state fluctuations are not affected during transport.

The statement of the extended fluctuation relations \cite{Bernard-Doyon-timeReversal} was obtained by extracting principles found in \cite{1751-8121-45-36-362001,Bernard2015} in the context of energy and charge transport in 1+1-dimensional CFT, and was argued to hold also in free particle models, later confirmed by various explicit calculations \cite{PhysRevLett.99.180601,DLSB,Y18}. This shows that the present formalism fully agrees with these results, and that, effectively, it generalises the method to nonlinear Euler hydrodynamics. As a consistency check, it is also a simple matter to see that the solution presented in \cite{Myers-Bhaseen-Harris-Doyon}, for interacting integrable models (and based on the present formalism), indeed reproduces the extended fluctuation relations when specialised to models without interactions.

\subsection{Fluctuations along rays and dynamical correlation functions of twist fields}\label{ssectray}

The proposal of subsection \ref{ssectmain} can be generalised to the statistics of the component of currents perpendicular to other space-time paths instead of the time-directed paths with constant space coordinates. Of particular interest is the application to dynamical correlation functions of twist fields, including order and disorder fields, in thermal states and other MES.

Consider, instead of $J^{(t)}$ defined in \eqref{Qt}, the quantity
\beq\label{Jell}
	J^{(\vec\ell)}=\int_0^1 \dd\vec{\ell}(s)\wedge \vec{\losymbol{j}}(\vec\ell(s)),
\eeq
determined by the path $\vec\ell = \{s\mapsto \vec{\ell}(s) = (\h x(s),\h t(s))\in\R^2:s\in[0,1]\}$, where $\dd\vec{\ell}(s) = (\dd \h x(s),\dd \h t(s))=\dd s\,(\dd \h x/\dd s,\dd\h t/ \dd s)$ is the infinitesimal tangent to the path, $\vec{\losymbol{j}} = (\losymbol{j},\losymbol{q})$ is the conserved current vector, and
\beq
	\dd\vec{\ell}\wedge \vec{\losymbol{j}} = \losymbol{j}\,\dd \h t - \losymbol{q}\,\dd \h x.
\eeq
Suppose (without loss of generality) that the path has end-points $\vec\ell(0) = (0,0)$ and $\vec\ell(1) = (x,t)$, and denote by $\ell = \sqrt{x^2+t^2}$ the Euclidean distance between the end-points. By current conservation, the quantity \eqref{Jell} is independent of the path chosen that connects $(0,0)$ to $(x,t)$. We may therefore choose it to be the segment of ray $x/t=\tan\theta$ (with $\theta\in[0,2\pi)$) determined by
\beq\label{ray}
	\vec\ell(s) = (sx,st) = s\ell\,(\sin\theta,\cos\theta).
\eeq
We are interested again in the large-$\ell$, scaled statistics, and so we must evaluate the expectation value
\beq\label{path}
	C_\lambda(x,t) = \bra \re^{\lambda J^{(\vec\ell)}}\ket_{\underline\beta}
\eeq
and the generator
\beq\label{Ftheta}
	F(\lambda;\theta) = \lim_{\ell\to\infty} \ell^{-1}\log C_\lambda(\ell\sin\theta,\ell\cos\theta).
\eeq
Extending the arguments presented in appendix \ref{appder}, or the simpler derivation of section \ref{ssectskewing}, it is a simple matter to derive, for the path \eqref{ray}, the flow (recall \eqref{istar})
\beq\label{flowbetaxi}
	\frc{\p}{\p\lambda} \mathsf{\beta}^i(\lambda;\theta) = -\Big(\sgn\big(\mathsf A(\lambda;\theta) - \tan\theta\, {\bf 1}\big)\Big)_{i_*}^{~i}
\eeq
which generalises \eqref{flowbeta} to the case $\theta\neq 0$. With this flow, the result takes a form that generalises \eqref{scgfres},
\beq\label{scgfresxi}
	F(\lambda;\theta) = \int_0^\lambda \dd\lambda'\,\big(\cos\theta\,\mathsf{j}(\lambda';\theta) - \sin\theta\, \mathsf{q}(\lambda';\theta)\big).
\eeq

Consider the limit $\theta\to\pi/2$, where the path is ``horizontal", lying on a the time slice $t=0$. In this case, the flow \eqref{flowbetaxi} does not depend on the flux Jacobian anymore, and is immediately solvable, simply effecting a shift of the Lagrange parameter $\beta^{i_*}$ proportional to $\lambda$, that is
\beq\label{flowbetaxipi2}
	\beta^i(\lambda;\pi/2) = \beta^{i}+\lambda\delta_{i_*}^i.
\eeq
From \eqref{scgfresxi} we have in this case $\p F(\lambda;\pi/2)/\p \lambda = -\mathsf{q}(\lambda;\pi/2)$, which, along with $F(0;\pi/2)=0$ and \eqref{flowbetaxipi2}, allows us to identify $F(\lambda;\pi/2)$ with a free energy difference. That is, we obtain
\beq\label{pi2}
	F(\lambda;\pi/2)= -\Delta f(\lambda)
\eeq
where $\Delta f(\lambda)$ is the specific (dimensionless) free energy difference
\beq
	\Delta f(\lambda) = f(\beta^\bullet + \lambda \delta_{i_*}^\bullet) - f(\underline\beta)
\eeq
with the specific free energy being $f(\underline\beta) = -\log Z(\underline\beta)$, where $Z(\underline\beta)$ is the partition function for Lagrange parameters $\underline\beta$.

The above results have perhaps their most interesting application to the evaluation of correlation functions of twist fields. Let us introduce the ``height fields" $\varphi(x,t)$ defined as $\losymbol{q}(x,t) = \p_x \varphi(x,t)$ and $\losymbol{j}(x,t) = -\p_t \varphi(x,t)$, which automatically solves the continuity relation \eqref{conslaw}. Differences of height fields $\varphi(x_2,t) - \varphi(x_1,t)$ count (for $x_2>x_1$, say) the quantity of charge present in $[x_1,x_2]$ at time $t$. Exponential of height fields $\Or(x,t) \propto e^{\lambda \varphi(x,t)}$ are a certain type of fields that have been studied in a variety of cases in the literature, and are referred to as {\em twist fields}\footnote{These are in fact twist fields associated with continuous symmetries, where the conservation law arise from the associated Noether current. More generally, twist fields may be associated with discrete symmetries as well, in which case however there is no Noether current, hence no obvious flux Jacobian.}. If the charge $Q$ is associated with an internal symmetry, then they are {\em local} in the general sense used in many-body quantum physics (they commute with the energy density at equal times). In particular, with $U(1)$ symmetry, the observable $e^{\lambda \varphi(x,t)}$ can be used to represent order parameters in many-body models: in the free Dirac fermion and in the Thirring model (or sine-Gordon model) it naturally occurs by bosonization and has applications to the transverse field Ising model and XXZ chains, see e.g. \cite{tsvelikbook,sachdevbook}. Certain classes of such twist fields can also be used to study entanglement entropy in free-particle models \cite{entropy}. Twist fields associated to {\em space-time symmetries} have also been studied recently \cite{conical}; they do not possess the conventional locality property of many-body physics anymore, although there is still path independence (a field $\varphi(x,t)$ can be defined independently of the path $\vec\ell$ chosen).

Interestingly, the SCGF \eqref{scgf} gives rise to the {\em leading exponential behaviour} of the dynamical two-point functions of twist fields:
\beq\label{twistfield}
	C_\lambda(x,t) = \bra e^{\lambda \varphi(0,0)}e^{-\lambda\varphi(x,t)}\ket_{\underline\beta} \asymp
	e^{\ell F(\lambda;\theta)}\qquad (x/t=\tan \theta,\;\ell=\sqrt{x^2+t^2}\to\infty)
\eeq
(recall footnote \ref{fn}). This is expected to hold in arbitrary maximal-entropy states of arbitrary many-body systems, integrable or not, including thermal states and, in integrable systems, GGEs. The evaluation of exponential behaviours of dynamical two-point correlation functions of order parameters -- which give a ``dynamical correlation lengths" -- is a notoriously difficult problem, for which there are only partial solutions (see e.g. \cite{sachdevbook}). Formula \eqref{twistfield} provides the first exact result in interacting models.

For equal-time, non-dynamical correlations ($\theta=\pi/2$), \eqref{twistfield} with \eqref{pi2} gives\footnote{BD acknowledges discussions with V. Alba at the Perimeter Institute, September 2017, that led to both \eqref{cfpi2} and \eqref{twistinhomo}.}
\beq\label{cfpi2}
	\bra e^{\lambda \varphi(0,0)}e^{-\lambda\varphi(x,0)}\ket_{\underline\beta}\asymp e^{-x\, \Delta f(\lambda)}
	\qquad (x\to\infty).
\eeq
In this case, some exact results already exist that confirm \eqref{cfpi2}. A specialisation to the $\Z_2$ twist fields of free Majorana fermions of \eqref{cfpi2} was proposed for arbitrary GGEs in \cite{ChenDoyon2014}, and shown to agree with results derived from special quantum quenches in the Ising model \cite{cef12}. In the context of entanglement entropy, taking into account the twist field interpretation of the R\'enyi entanglement entropy as expressed in \cite{entropy}, the exact results of \cite{ac-16,ac-17c,AlbaCalabrese2017i,AlbaCalabrese2017ii,MAC2018} can be interpreted as giving the leading exponential decay of permutation twist fields correlation functions in GGEs of interacting integrable systems, which also agree with\footnote{Results \eqref{cfpi2} and \eqref{twistinhomo} hold as well for twist fields that are associated with discrete symmetries, as $\Delta f(\lambda)$ is well defined in these cases also. Hence the results are applicable to the study of the entanglement entropy in interacting models. They give an exact formula in integrable models using the thermodynamic Bethe ansatz \cite{ZAMOLODCHIKOV1990695}: in a state described by the source term $w(\theta)$, for the $n$-copy permutation twist-field two-point function $\bra{\cal T}_n(0,0)\b{\cal T}_n(x,0)\ket$, we have $\Delta f_n = \int \dd p(\theta)\,\log\lt(\frc{1+\re^{-\ep_n(\theta)}}{(1+\re^{-\ep_1(\theta)})^n}\rt)$, with pseudoenergy $\ep_n(\theta)$ having source term $nw(\theta)$ implementing the $n$-times larger imaginary time direction induced by the twist property, and $p(\theta)$ being the momentum function.} \eqref{cfpi2}.

\begin{rema} From the path independence of the variable \eqref{Jell}, we may choose different paths than the ray \eqref{ray} and try to evaluate the related flow. For instance, we may choose the piecewise straight path which goes first in the time direction $(\h x(s),\h t(s)) = (0,2st),\;s\in[0,1/2]$, and then in the space direction $(\h x(s),\h t(s)) = ((2s-1)x,t),\;s\in[1/2,1]$. Evaluating the flow equation from this is, however, more complicated. It might be tempting to think that the result will separate into two contributions, one from each straight piece, $F(\lambda;\theta)$ being proportional to a sum of $F(\lambda;0) = F(\lambda)$ (the SCGF calculated in subsection \ref{ssectmain}) and $F(\lambda;\pi/2) = -\Delta f(\lambda)$ (the difference of free energy densities). However, this is generically incorrect: for instance, because of ballistic transport, the space integral $\int _0^x\,\dd \h x\,\bra \losymbol{q}(\h x,t),\losymbol{q}_i(0,0)\ket_{\underline\beta}^{\rm c}$ does not necessarily vanish in the limit $x,t\to\infty$. Likewise, the evaluation of $F(\lambda;\theta)$ for any choice of path that is not straight (at the scale set by $\ell\to\infty$) may receive contributions from correlations between separated portions of the path if they are connected by ballistic transport of normal modes. See subsection \ref{ssectbreaking}.
\end{rema}

\begin{rema} It is a simple matter to generalise \eqref{pi2} and \eqref{cfpi2} to states which are inhomogeneous at Euler scales, and described by fluid cells with position-dependent Lagrange parameters $\underline\beta(\h x)$. By using the idea of local entropy maximisation, one simply expects each ``fluid cell" at $\h x$ to produce a contribution proportional to its free energy difference $\Delta f_{\h x}(\lambda)$, and thus
\beq\label{twistinhomo}
	\bra e^{\lambda \varphi(0,0)}e^{-\lambda\varphi(x,0)}\ket^{\rm c}_{\{\h x\mapsto \underline\beta(\h x)\}}\asymp e^{-\int_0^x \dd \h x\, \Delta f_{\h x}(\lambda)}.
\eeq
An ansatz of this form appeared, for the order parameter correlation functions in the Ising model, in \cite{bc-1}, and was verified against direct numerical calculations. This is also connected to formulae for entanglement entropies in inhomogeneous states \cite{bfpc-18}. Again, this is expected to be valid for arbitrary twist fields and in interacting models as well. However, for dynamical correlation functions in inhomogeneous, non-stationary states, because of correlations produced by ballistically propagating modes, we do not expect the simple generalisation of \eqref{scgfresxi} and \eqref{twistfield} to similar integrals over space-time paths to be correct. The theory developed in \cite{doyoncorrelations} for charge-density dynamical correlations in inhomogeneous, non-stationary states might be useful for solving this problem.
\end{rema}

\subsection{Divergence of scaled cumulants and non-Gaussianity}\label{ssectbreaking}

We now explain in what situations the limit defining the scaled cumulants in the SCGF \eqref{scgf} may be divergent (in which case the large deviation principle expressed in \eqref{ld} fails), and what the meaning of this may be.

We first note that the scaled cumulants of \eqref{scgf} are time-integrated, connected, multi-point correlation functions of local current observables, as expressed in \eqref{cumul}. They exist if correlation functions of local currents cluster fast enough at large time separations. More generally, the cumulants on arbitrary rays, \eqref{path} and \eqref{Ftheta} defined in subsection \ref{ssectray}, exist if correlation functions cluster fast enough at large separations along rays in space-time. This is the main assumption behind the results we have presented: that of strong enough clustering. Clustering in space can be shown on quite general grounds in Gibbs states of local hamiltonians in one dimension \cite{Araki} (in higher dimensions it may fail as thermal phase transitions are possible). However, clustering along nontrivial rays is more subtle. In what situations may it be broken?

Consider for instance the current-density correlation. In the Euler scaling limit (see appendix \ref{appder}), it can be expressed (formally) as \eqref{jqcorr}. It is clear that this is zero at $x/t=\xi$ if there is no effective velocity $v^{\rm eff}_i$ that takes the value $\xi$. This means that, far along this ray, the current-density connected correlation function is expected to vanish exponentially fast. Likewise, current-current correlation functions,
\beq\label{jjcorr}
	\bra \losymbol{j}_i(x,t),\losymbol{j}_j(0,0)\ket^{\rm c}_{\underline\beta} \sim (\mathsf A^2\delta(x-\mathsf At)\mathsf C)_{ij},
\eeq
vanish exponentially fast if $v^{\rm eff}_i\neq \xi\;\forall \;i$. In fact, from the strong hydrodynamic projection principle \eqref{proj}, and the slightly stronger version \cite[Eq 3.35]{doyoncorrelations}, this holds more generally for local observables. However, if there is an effective velocity at the value $\xi$, then the Euler-scale expression diverges. This is generically associated with algebraic instead of exponential clustering of correlation functions.

There is therefore a link between the strength of the correlation on the ray $\xi$ and the presence or not of an effective velocity with value $\xi$. Recall that the effective velocity is the velocity of ballistic propagation of the normal modes of the fluid (such as pressure waves in air -- sound waves). It is physically natural that ballistically propagating normal modes create strong correlations along their paths. These are sometimes referred to as ``sound peaks", or ``heat peaks". Such strong correlation also occur naturally in rarefaction waves \cite{BressanNotes}: there, the state at ray $\xi$ is such that  $v^{\rm eff}_j=\xi$ for some $j$.

As a consequence, the large deviation principle \eqref{ld} does not hold if there exists such a normal mode propagating along the ray -- a ``co-propagating mode".  In fact it is possible to argue from our explicit results that this is the case. The calculation of cumulants requires us to take $\lambda$ derivatives. Consider the case $\xi=0$ for simplicity. Clearly, from \eqref{scgfres}, the first cumulant (the average current) is expected to be finite. From \eqref{flowbeta}, the second cumulant also is expected to be finite, although its exact value is ambiguous (in the present theory) if there is a co-propagating mode, because of the ambiguity of the sign function at 0. However, for the third cumulant, we need to take another derivative, which does not exist if there is such a co-propagating mode that couples to the current of interest (i.e. such that the corresponding element of the  $\mathsf{A}$ is nonzero), because of the discontinuity of the sign function. In this case, all scaled cumulants $c_n$ for $n\geq 3$ do not exist. Because a discontinuity points to a diverging derivative, this indicates that these higher cumulants are actually divergent.

The divergence of scaled cumulants may be interpreted as a ``dynamical phase transition", a concept widely studied in the nonequilibrium large-deviation theory of stochastic models, see for instance \cite{Bodineau2005,Garragan2007,Garragan2009,Espigares2013,Jack2014,Hurtado2014,Tsobgni2016,Tizon2017,Lazarescu2017}. Indeed, we see a change in the fluctuation spectrum for a dynamical quantity as a parameter of the state is modified, or as the bias $\lambda$ is modified. On both sides of the phase transition point, the cumulants will take different forms, as $\sgn(A)$ changes discontinuously. \newline\indent
One subtle point is worth mentioning. It seems as though in order for this phase transition to occur, the co-propagating mode should be ``isolated". We define a {\em non-}isolated mode as a mode whose effective velocity is part of a continuum of effective velocities, in a state where the associated modes are smoothly populated and smoothly coupled to the charge whose transport we study. We claim that if a mode is non-isolated, then no discontinuity appears due to the ensuing smoothing, and no divergence emerges.  This is what we observe for transport of generic local conserved quantities in integrable systems \cite{Myers-Bhaseen-Harris-Doyon}, where there is a continuum of quasi-particle velocities. There are situations where isolated effective velocities may be present in integrable systems, for instance for spin transport in the XXZ spin chain, see \cite{PirolietalNonBall}, and it might be possible to study transport of charges that couple to a single quasi-particle velocity.\newline\indent
What does the divergence of scaled cumulants mean? We propose that the leading Gaussian form of the fluctuation spectrum may be broken in states with a co-propagating mode. Recall that the large-deviation principle -- which says that all cumulants scale with $t$ in \eqref{scgf} -- is an extension of the law of large numbers: subtracting the average $t\b\j$, the fluctuations of $(J^{(t)}-t\b\j)/\sqrt{t}$ are Gaussian at large $t$, with nonzero second cumulant and vanishing higher cumulants (as the higher cumulants of this variable receive a scaling $t^{-n/2}$ instead of the $t^{-1}$ used in \eqref{scgf}). The divergence of the scaled cumulants of large-deviation theory $c_n,\;n\geq 3$ suggest that the cumulants of $(J^{(t)}-t\b\j)/\sqrt{t}$ might {\em no longer be vanishing}, thus breaking Gaussianity.\newline\indent
Crucially, this may have a connection with {\em nonlinear fluctuating hydrodynamics} \cite{SpohnNonlinear,1742-5468-2015-3-P03007,ChenDeGierHirikiSasamotoFluctuHydro18}. Nonlinear fluctuating hydrodynamics can be used to describe the broadening of correlation peaks occurring along the ballistic rays of normal modes, and correlations in rarefactions waves. It is observed that the first order (linear) expansion of noisy hydrodynamic equations, leading to Gaussian fluctuations, vanishes if there is a co-propagating mode, and the next order needs to be taken, leading to fluctuations in the KPZ class. Nonlinear fluctuating hydrodynamics describes more than the particular current fluctuations studied here. Nevertheless, the present theory is in agreement with it, and provides an additional confirmation of the breaking of Gaussianity. In particular, with \cite{Myers-Bhaseen-Harris-Doyon} and the above discussion, this shows that, as already predicted in \cite{SpohnNonlinear}, {\em the KPZ fluctuations of nonlinear fluctuating hydrodynamics do not generically occur in integrable systems, except perhaps for very specific variables and states}, such as the spin in spin transport problems of the XXZ chain (as \cite{PirolietalNonBall} indicates that there is an isolated mode), or perhaps observables able to isolate quasi-particle velocities. The present theory does not yet confirm the KPZ class of fluctuations predicted by nonlinear fluctuating hydrodynamics. We hope to develop these ideas in a future work.

\section{Application to conformal hydrodynamics in arbitrary dimensions}
\label{sectCFT}

The proposal of subsection \ref{ssectmain} applies to all systems in the class described in subsection \ref{ssectmes}. Although the focus was on one dimension, the formalism applies as well to effectively one-dimensional setups in higher-dimensional systems. The goal of this section is to provide a non-integrable example of the formalism, and to show how dimensional reduction is performed.

There are many non-integrable systems which admit ballistic transport. One family of examples is relativistic or Galilean quantum and classical field theory, in arbitrary dimension. Higher-dimensional relativistic conformal field theory is particularly interesting, as it makes predictions for quantum systems tuned to quantum critical points at small but nonzero temperatures. In dimensions higher than one, very few results are available. Further, the equations of state -- giving conformal hydrodynamics -- are almost completely determined by the symmetries, making the present formalism immediately applicable.

Here we study the important example of energy transport in conformal hydrodynamics of arbitrary dimension, obtaining explicitly the flow \eqref{flowbeta}, expressions for the cumulants $c_2$, $c_3$ and $c_4$, and numerically evaluating the SCGF $F(\lambda)$. All these are, to our knowledge, new results\footnote{The cumulant $c_2$, as mentioned, follows from a general formula for current-current sum rule that was already known \cite{1742-5468-2015-3-P03007,SciPostPhys.3.6.039}; however it was never worked out explicitly in conformal hydrodynamics.}. We further note that conformal hydrodynamics was used in order to obtain exact nonequilibrium steady states of quantum critical systems in the partitioning protocol \cite{BDLS,CKY,LSDB,SH}. The properties of $F(\lambda)$ in such states will be analysed in a future publication.

\subsection{Reduction to a one-dimensional hydrodynamic problem}

Consider conformal field theory in $d> 1$ dimensions of space, and recall that the energy-momentum tensor $\losymbol T^{\mu\nu}$ satisfies $\losymbol T^{\mu\nu} = \losymbol T^{\nu\mu}$ (Lorentz invariance), $\losymbol T^\mu_{\ \mu}=0$ (scale invariance) and $\p_\mu \losymbol T^{\mu\nu} = 0$ (conservation of energy and momentum). Assume the system not to be integrable -- this is the generic situation, and excludes free field theory. Then the full MES manifold is that of boosted thermal states. For simplicity, consider transport in the direction $x^1$ (with $x^0$ the time coordinate), and the associated momentum operator $P = \int \dd^{d} x\,\losymbol T^{01}(x)$. In this case, we may restrict to the space of thermal states boosted in that direction \cite{BDLS}, with density matrices
\beq\label{higherdstate}
	\re^{-\beta^1 H - \beta^2 P} = \re^{-\beta_\rre(\cosh\theta\, H - \sinh \theta\, P)}
\eeq
and corresponding state denoted by $\bra \cdots\ket^{(d)}_{\beta_1,\beta_2}$. Here $H = \int \dd^{d} x\,\losymbol T^{00}(x)$ is the Hamiltonian, the rest-frame temperature is $T_\rre=\beta_\rre^{-1}$, and the Lorentz boost is of rapidity $\theta$. By relativistic and conformal invariance, expectation values of energy-momentum components take the form
\al{\label{Tav}
\bra \losymbol T^{\mu \nu}(x,t)\ket^{(d)}_{\beta_1,\beta_2} = a T_\rre^{d+1}((d+1)u^{\mu} u^{\nu} + \eta ^{\mu \nu}) \, , \quad  \eta ^{\mu \nu} = \text{diag}(-1,1,1...1)
} 
where $u^\mu = (\cosh\theta, \sinh\theta, 0,...,0)^\mu$ and $a$ is a model-dependent positive constant.

In order to make the connection with the formalism developed, we need to render the system effectively one-dimensional. This can be done by integration over the transverse directions. Specifically, we assume the transverse space $S_\perp$, with coordinates $x^\perp = (x^2,\ldots,x^d)$, to be of $d-1$-dimensional hyperarea $V_\perp$ and to be periodic in all its coordinates, with equal periods. Let us denote by $\bra \cdots\ket^{(d,V_\perp)}_{\beta_1,\beta_2}$ the resulting state with density matrix of the form \eqref{higherdstate}; in particular, the limit of infinite transverse hyperarea reproduces the infinite-volume results,
\beq\label{infvol}
	\lim_{V_\perp\to\infty} \bra \losymbol T^{\mu\nu}(x,t)\ket_{\beta_1,\beta_2}^{(d,V_\perp)} = \bra \losymbol T^{\mu\nu}(x,t)\ket_{\beta_1,\beta_2}^{(d)}.
\eeq
We then define one-dimensional densities as
\beqa
	\losymbol{q}_1^{(V_\perp)}(x,t) &=& \int_{S_\perp} \dd^{d-1} x^\perp\, \losymbol T^{00}(x,x^\perp,t) \n
	\losymbol{j}_1^{(V_\perp)}(x,t) = \losymbol{q}_2^{(V_\perp)}(x,t) &=& \int_{S_\perp} \dd^{d-1} x^\perp\,\losymbol T^{01}(x,x^\perp,t) \n
	\losymbol{j}_2^{(V_\perp)}(x,t) &=& \int_{S_\perp} \dd^{d-1} x^\perp\, \losymbol T^{11}(x,x^\perp,t)
\eeqa
(the equality $\losymbol{j}_1^{(V_\perp)}(x,t) = \losymbol{q}_2^{(V_\perp)}(x,t)$ comes from $\losymbol T^{01}(x,x^\perp,t) = \losymbol T^{10}(x,x^\perp,t)$, due to Lorentz invariance). It is a simple matter to see that
\beq
	\p_t \losymbol{q}_i^{(V_\perp)}(x,t) + \p_x \losymbol{j}_i^{(V_\perp)}(x,t) = 0,\qquad i=1,2,
\eeq
and correlation functions cluster at large longitudinal distances. Therefore this is an effectively one-dimensional system, with two conserved charges, and assuming that the transverse direction does not give rise to additional thermodynamic degrees of freedom (this in particular assumes no turbulent instabilities, see the discussion in \cite{BDLS}), one-dimensional Euler hydrodynamics apply. We may then ask about the SCGF for the energy current $\losymbol{j}_1^{(V_\perp)}$ as defined in \eqref{scgf}, and the general discussion and results of section \ref{sectmain} hold. In fact, it is convenient to divide the SCGF by $V_\perp$, and so we consider
\beq
	F^{(V_\perp)}(\lambda) = \lim_{t\to\infty} (tV_\perp)^{-1}\log \big\bra \re^{\lambda \int_0^t \dd s \int_{S_\perp} \dd^{d-1} x^\perp \losymbol T^{01}(t,0,x^\perp)}\big\ket_{\beta_1,\beta_2}^{(d,V_\perp)}.
\eeq

By clustering in $d+1$-dimensional space-time, all cumulants generated by $F(\lambda)$ have a finite limit as $V_\perp\to\infty$, and we look for
\beq\label{FlargeV}
	F(\lambda) = \lim_{V_\perp\to\infty} F^{(V_\perp)}(\lambda).
\eeq
Clearly, this is obtained by taking the large-$V_\perp$ limit of the solution presented in subsection \ref{ssectmain} for the effectively one-dimensional system; that is, of the expression \eqref{scgfres} divided by $V_\perp$, with \eqref{flowbeta}. It is then sufficient to know the following large-$V_\perp$ limits of conserved densities and currents:
\beq\label{largeV}
	\mathsf{q}_i,\;\mathsf{j}_i:=\lim_{V_\perp\to\infty}V_\perp^{-1}\bra\losymbol{q}_i^{(V_\perp)}(x,t)\ket_{\beta_1,\beta_2}^{(d,V_\perp)} ,\;
	\lim_{V_\perp\to\infty} V_\perp^{-1}\bra\losymbol{j}_i^{(V_\perp)}(x,t)\ket_{\beta_1,\beta_2}^{(d,V_\perp)} 
\eeq
and to solve for the flow \eqref{flowbeta} with the flux Jacobian given by
\beq
	\mathsf A_i^{~j} = \frc{\p \mathsf{j}_i}{\p\mathsf{q}_j}.
\eeq
Using \eqref{infvol} as well as homogeneity in the transverse direction, the limits in \eqref{largeV} are given exactly by the expression on the right-hand side of \eqref{Tav} for $\mu,\nu\in\{0,1\}$. Thus we have fully reduced the problem of evaluating \eqref{FlargeV} to a one-dimensional hydrodynamic problem.

\subsection{Exact SCGF and cumulants}

Using \eqref{Tav}, we have more explicitly
\al{\label{q1}
\mathsf{q}_1 &= aT^{d+1}_\rre (d \cosh^2\theta +\sinh^2\theta)\\
\label{j1}
\mathsf{j}_1 = \mathsf{q}_2 &= a(d+1)T^{d+1}_\rre\cosh\theta\sinh\theta\\
\label{j2}
\mathsf{j}_2 &=  aT^{d+1}_\rre(\cosh^2\theta + d\sinh^2\theta).
}
The flux Jacobian takes the form
\al{\label{dCFT A}
\mathsf A_{i}^{~j} =  \frc{\p\mathsf{j}_i}{\p{\mathsf{q}_j}} = \frc1{d\cosh^2\theta-\sinh^2\theta} \begin{pmatrix}
0 &&d\cosh^2\theta-\sinh^2\theta \\ \displaystyle \cosh^2\theta-d\sinh^2\theta
&& \displaystyle (d-1)\sinh2\theta
\end{pmatrix}_{ij}.
}
As the flux Jacobian is not state-independent if $d>1$, the extended fluctuation relations do not hold in higher-dimensional CFT\footnote{This invalidates the conjecture made in \cite{BDLS} for the SCGF, which was based on the extended fluctuation relations.}. The matrix $\sgn(\mathsf A)$ is obtained by diagonalising $\mathsf A$, and taking the sign of the eigenvalues. The eigenvalues of $\mathsf A$ are:
\al{\label{xipm}
v^{\rm eff}_\pm 
= \frc{d-1}{2(d\cosh^2\theta-\sinh^2\theta)}\lt(\sinh2\theta \pm \alpha\rt)\quad\mbox{where}\quad \alpha = \frc{2\sqrt{d}}{d-1} = 2 \frac{v_s}{1 - v_s^2}
}
with the speed of sound of conformal hydrodynamics given by $v_{\rm s} = 1/{\sqrt{d}}$. Clearly, $\sgn(v^{\rm eff}_\pm) = \sgn(\sinh2\theta \pm\alpha) = \sgn(\theta\pm\theta_{\rm s})$ with the sound rapidity $\theta_{\rm s}$ defined by $\tanh\theta_{\rm s} = v_{\rm s}$. Define $  \xi_1 =  \sgn{v^{\rm eff}_+} + \sgn{v^{\rm eff}_-}$ and $ \xi_2 =  \sgn{v^{\rm eff}_+} -  \sgn{v^{\rm eff}_-}$. They take the following values:
\beq\label{explicit xi1 xi2}
	\xi_1 = \lt\{\ba{ll} 2 \sgn(\theta), & |\theta|>\theta_{\rm s} \\
	0, & |\theta|<\theta_{\rm s}
	\ea\rt.,\quad
	\xi_2 = \lt\{\ba{ll} 0, & |\theta|>\theta_{\rm s} \\
	2, & |\theta|<\theta_{\rm s}.
	\ea\rt.
\eeq
That is, $\xi_1$ is nonzero for supersonic rapidities and zero otherwise, and $\xi_2$ is nonzero for infrasonic rapidities and zero otherwise. We then have
\al{
\sgn{\mathsf A} = \frac{1}{2\alpha}\begin{pmatrix}
 \xi_1\alpha -  \xi_2  \sinh 2\theta && 2 \xi_2 (\cosh^2 \theta + \gamma) \\
 -2 \xi_2  (\sinh^2\theta - \gamma) && \xi_1\alpha +\xi_2  \sinh 2\theta
\end{pmatrix}\quad\mbox{where}\quad \gamma = \frac{1}{d-1} = \frc{v_{\rm s}^2}{1-v_{\rm s}^2}.
}

Consider the SCGF for energy transport. We fix  $i_{*}$ = 1 in \eqref{flowbeta} and let the Lagrange multipliers become $\lambda$ dependent. Then using $\beta_1 = \beta_\rre \cosh\theta $ and  $\beta_2 =- \beta_\rre \sinh\theta$, we obtain
\al{
\p_{\lambda}
\beta_\rre(\lambda)  & = -\frac{1}{2 \sqrt{d}}
\Big( \xi _2(\lambda ) \sinh (\theta (\lambda ))+ \sqrt{d}\xi _1(\lambda ) \cosh (\theta (\lambda )) \Big) \n
\p_\lambda
\theta(\lambda) & =  \frac{1}{2 \beta_\rre(\lambda)} \Big( \xi _1(\lambda ) \sinh (\theta (\lambda ))+\sqrt{d} \xi _2(\lambda ) \cosh (\theta (\lambda )) \Big).
\label{flow in higher d CFT}
}
We may evaluate the SCGF according to \eqref{scgfres} by integrating the current \eqref{j1}
\beq\label{jlambda}
	\mathsf{j}_1(\lambda) = a \frc{d+1}2 T_{\rm rest}^{d+1}(\lambda) \sinh2\theta(\lambda).
\eeq
Cumulants are simply obtained by taking derivatives with respect to $\lambda$ and setting $\lambda=0$.

The derivatives may be readily evaluated:
\al{
\p_{\lambda} \losymbol{j}_1(\lambda) = a(d+1)T_{\rm rest}^{d+1}(\lambda) \rndb{\frac{d+1}{2}\sinh2 \theta(\lambda)\,  \p_{\lambda} \log T_\rre(\lambda)  + \cosh 2 \theta (\lambda) \,\p_{\lambda} \theta(\lambda)}. 
}
Reading off the required identities from \eqref{flow in higher d CFT}, setting $\lambda = 0$ and using the explicit form of $\xi_{1/2}$ given by \eqref{explicit xi1 xi2}, we obtain the second cumulant:
%
\al{\label{cumuc2}
c_2 = \frac{a(d+1)T_{\rm rest}^{d+2}}{{2 \sqrt{d}}} \times \lt\{\ba{ll}  \sqrt{d} \sinh |\theta |\, ((d+3) \cosh (2 \theta )+d+1), & |\theta|>\theta_{\rm s} \\
	 \cosh (\theta )\, ((3 d+1) \cosh (2 \theta )-(d+1)), & |\theta|<\theta_{\rm s}
	\ea\rt.
}
where $\theta = \theta(0)$ and  $\beta_\rre = \beta_\rre(0)$. This process easily generates cumulants of $n^{\text{th}}$ order. As an example, using the Mathematica symbolic calculation software, we calculate $c_3$,
\al{\label{cumuc3}
c_3 =\frac{a(d+1) T_{\rm rest}^{d+3}}{4 d } \times \lt\{\ba{ll} d(d+3) \sinh (2 \theta ) \,((d+5) \cosh (2 \theta )+d-1), & |\theta|>\theta_{\rm s} \\
	2 (2 d+1) \sinh (2 \theta ) \,((3 d+1) \cosh (2 \theta )+d-1), & |\theta| < \theta_{\rm s}.
	\ea\rt.
}
and $c_4$:
\al{\label{cumuc4}
c_4 =&  \frac{a(d+1)  T_\rre^{d+4}}{8 d^{3/2} } \times \nonumber \\
& \,   \lt\{\ba{ll}  d^{3/2} (d+3) \sinh |\theta|\,\times &\\((d+5) (d+7) \cosh (4 \theta )+ 4 (d+1)(d+5) \cosh (2 \theta )+ 3 (d+1)(d+3))  , & |\theta|>\theta_{\rm s} \\
	 2 (2 d+1)  \cosh (\theta )\times&\\
	 ((3 d+1) (5 d+3) \cosh (4 \theta )-4 (5 d+3) \cosh (2 \theta )+(d+3)^2) . & |\theta|<\theta_{\rm s}.
	\ea\rt.
}

\begin{figure}[h!]
    \centering
    \subfloat[SCGF: $d=2$]{{\includegraphics[scale=0.3]{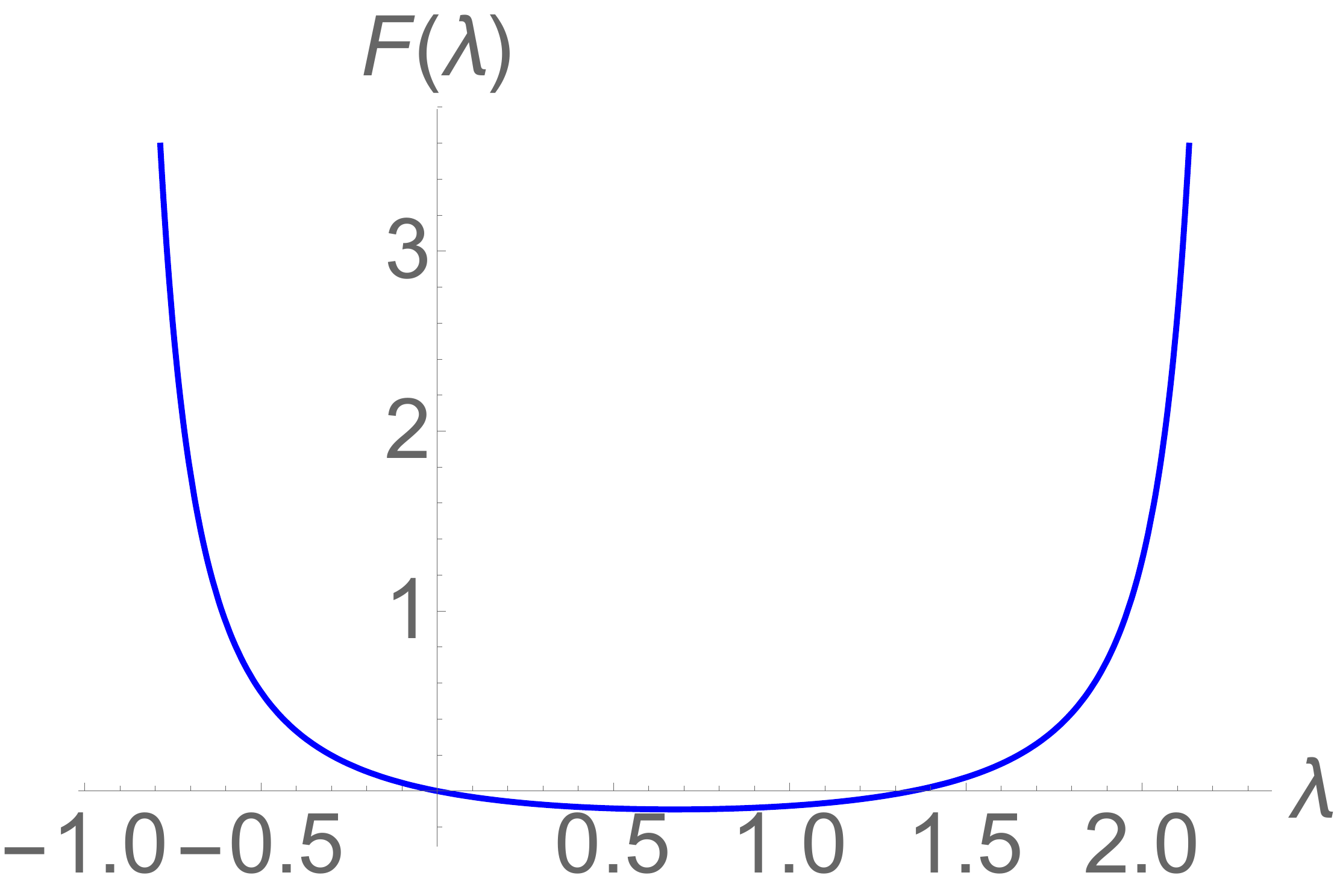} }}%
    \qquad
    \subfloat[SCGF: $d=3$]{{\includegraphics[scale=0.3]{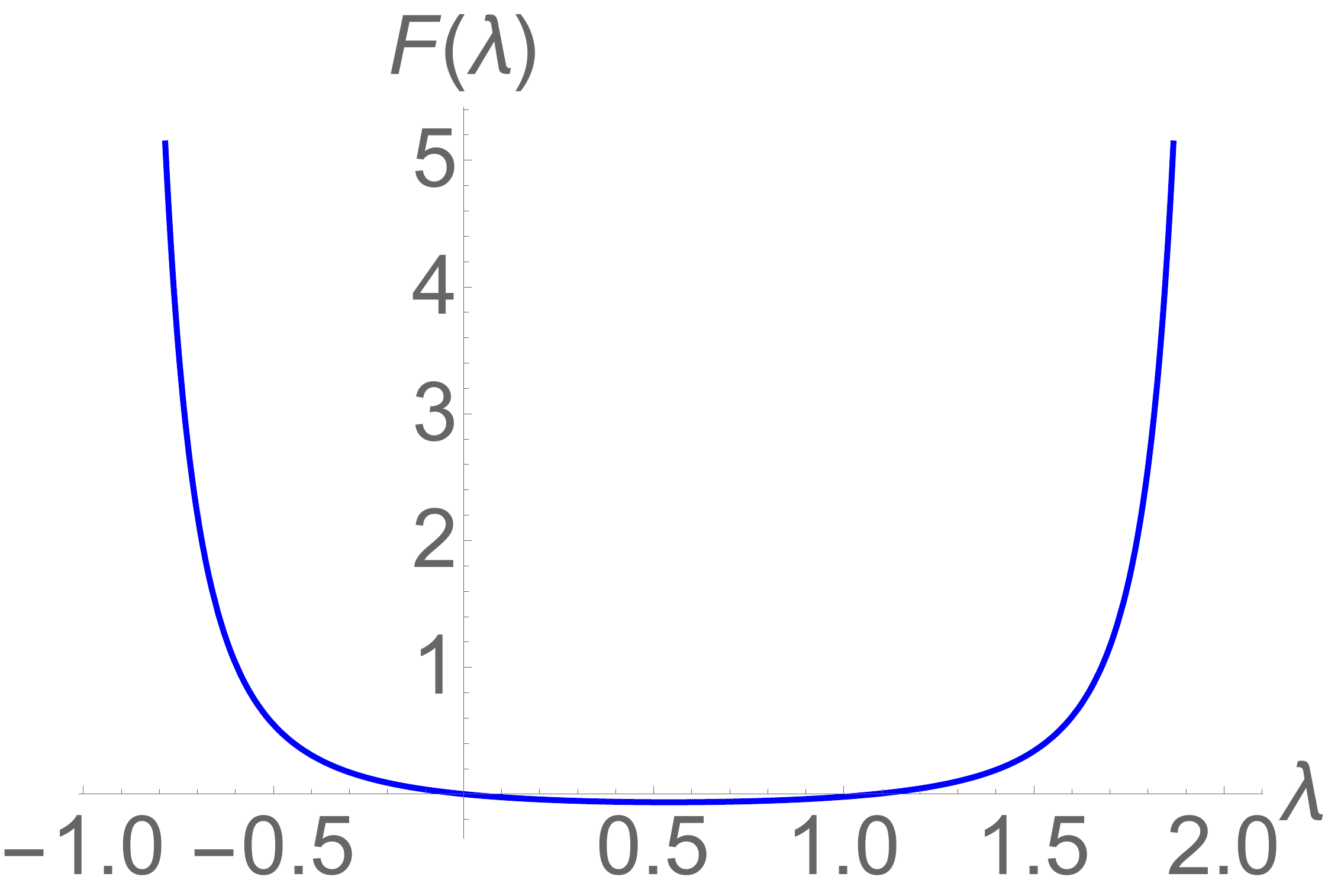} }}%
    \caption{Figure showing numerical solutions for the SCGF of a CFT in 2D (a) and 3D (b). Using $\beta_\rre (\lambda = 0)  = 1.73145,  \theta (\lambda = 0) = -0.55 $}%
    \label{fig:SCGF d 2 and 3}%
\end{figure}

One can determine the SCGF itself by numerically solving \eqref{flow in higher d CFT} to find $\beta_\rre(\lambda)$ and $\theta(\lambda)$, then inserting these expressions into \eqref{jlambda}, and numerically integrating \eqref{scgfres} to obtain $F(\lambda)$. We show the results for $d=2$ and $d=3$ in Fig.~\ref{fig:SCGF d 2 and 3}. We verify that the resulting functions are convex, as they should by the general theory \cite{BressanNotes}. From these plots many insights can be drawn out. In particular it is possible to verify that the powerful fluctuation relations of Gallavotti-Cohen type \cite{PhysRevLett.74.2694,PhysRevLett.78.2690,PhysRevE.56.5018,Crooks98,0305-4470-31-16-003,Lebowitz1999,PhysRevLett.92.230602,Bernard-Doyon-timeReversal} hold. We leave an in-depth analysis to a future work.


Finally, we note that the phenomenon discussed in subsection \ref{ssectbreaking} can be explicitly seen here.  Consider a thermal state boosted the sound velocity, $\theta = \pm\theta_{\rm s}$. In this case, as a consequence of the discontinuities in $\xi_{1,2}$ in \eqref{explicit xi1 xi2}, the derivatives of $\beta_{\rm rest}(\lambda)$ and $\theta(\lambda)$ with respect to $\lambda$ have discontinuities at $\lambda=0$. This implies that the third derivative of $F(\lambda)$ does  not exist at $\lambda=0$. That is, the scaled cumulants $c_n$ for $n\geq 3$ do not exist, and the large-deviation principle is broken; these higher cumulants are expected to be divergent. Intuitively, when an object moves in a medium exactly at the speed of sound, there is a build up of linear waves generated. At the macroscopic scale, this appears to increase correlations of transported energy to such an extent so as to modify the scaling of higher-order cumulants with time. There thus appear to be  a dynamical phase transition. On both sides of the phase transition point, the cumulants take different form, as is clear from \eqref{cumuc2} and \eqref{cumuc3}.

\section{Conclusion}\label{sectconclu}

In this paper, we show how to calculate the scaled cumulant generating function (or full counting statistics) for transport of any conserved quantity in stationary, homogeneous, clustering states of many-body systems, in or out of equilibrium. The technique is based on large-deviation theory and the result is expressed in terms of quantities readily available from the Euler hydrodynamics description of the system. This can be seen as a nonlinear generalisation of the construction in \cite{1751-8121-45-36-362001,Bernard2015} for 1+1-dimensional conformal field theory to interacting integrable \cite{Myers-Bhaseen-Harris-Doyon} and non-integrable models. We show that the extended fluctuation relations proposed in \cite{Bernard-Doyon-timeReversal} hold whenever the Euler hydrodynamics is linear. We extend the theory to arbitrary rays and make the connection with spacio-temporal correlation functions of twist fields, which have applications to order-parameter correlations. We also explain in what situations the theory may break and ``dynamical phase transitions" may occur, making connection with nonlinear fluctuating hydrodynamics. Finally, we give the example of conformal hydrodynamics in arbitrary dimensions, obtaining the first exact results for energy transport cumulants in spatial dimensionality higher than 1. We observe, in this example, a breaking in thermal states boosted at the sound velocity.

Future works would include an in-depth study of fluctuations in higher-dimensional conformal hydrodynamics, especially in the nonequilibrium steady states constructed in \cite{BDLS,CKY,LSDB,SH}, as well as the analysis in other non-integrable models where ballistic transport exist, such as anharmonic chains or one-dimensional hard rods with alternating masses \cite{BrunetDerridaGerschenfeld10}. An understanding of how the Gallavotti-Cohen-type fluctuation relations emerge from our theory in the general setting of the Riemann problem of Euler hydrodynamics is also lacking. It would be interesting to extend the ideas developed here to include diffusion, and to make potential connections with macroscopic fluctuation theory. A more in-depth study of the exact formulae for dynamical correlation functions for order parameters and other twist-field in stationary, homogeneous states would also be needed. A full connection with nonlinear fluctuating hydrodynamics, and fluctuations in the KPZ class, would be very interesting. It would be nice to see if the formalism can be appropriately extended in order to include the exact logarithmic large-deviation results found recently \cite{Moriya2019} (based on \cite{Eisler2013}).

Finally, as we already remarked, the extended fluctuation relations form a marker of ``freeness" -- they hold in free-particle models and 1+1-dimensional conformal field theory. In these cases, Euler hydrodynamics is linear --  equivalently, the natural flux Jacobian is state-independent -- and fluctuations in transport are directly related to initial-state fluctuations. This is in contrast to nonlinear Euler hydrodynamic systems, where the nonlinear evolution affects the structure of transport fluctuations as per the theory developed here. Hydrodynamic diffusion has also been argued to be a signal for the lack of interactions \cite{SpohnInteracting}, which has been confirmed for integrable models \cite{dNBD,dNBD2}, and we observe that all known models where extended fluctuation relations hold also have vanishing hydrodynamic diffusion, and vice versa. Is there a relation between state-independence of the natural flux Jacobian, and the vanishing of the diffusion matrix (although these two objects operate at different hydrodynamic scales)? Interestingly, this potential relation is further brought to light by a recent result  \cite{GHKV2018}, which can be interpreted as connecting, in the context of integrable systems, state differentiation of the flux Jacobian to the diagonal elements of the diffusion matrix. Why these two different hydrodynamic scales may be connected in this way remains to be explained.

\medskip

{\bf Acknowledgments.} We thank M. Joe Bhaseen, Denis Bernard, Juan P. Garrahan, Rosemary J. Harris, Tomohiro Sasamoto, Herbert Spohn, Pierpaolo Vivo, Takato Yoshimura and Claudio Zeni, and concerning subsection \ref{ssectray}, Vincenzo Alba, Pasquale Calabree and Mart\'on Kormos, for discussions, comments and suggestions. BD acknowledges funding from the Royal Society under a Leverhulme Trust Senior Research Fellowship, ``Emergent hydrodynamics in integrable systems: non-equilibrium theory", ref. SRF\textbackslash R1\textbackslash 180103, and from the EPSRC through a standard grant,  ``€œEntanglement Measures, Twist Fields, and Partition Functions in Quantum Field Theory", ref.~EP/P006132/1. JM acknowledges  funding  from  the  EPSRC Centre  for  Doctoral  Training  in  Cross-Disciplinary  Approaches  to  Non-Equilibrium  Systems  (CANES)  under grant EP/L015854/1. BD is grateful to \'Ecole Normale Sup\'erieure de Paris for an invited professorship (2018), and acknowledges hospitality and funding from the Erwin Schr\"odinger Institute in Vienna (Austria) during the program ``Quantum Paths" (2018), and the Galileo Galilei Institute in Florence (Italy) during the program ``Entanglement in Quantum Systems" (2018).  BD's research was supported in part by Perimeter Institute for Theoretical Physics. Research at Perimeter Institute is supported by the Government of Canada through the Department of Innovation, Science and Economic Development and by the Province of Ontario through the Ministry of Research, Innovation and Science. Both authors thank the Centre for Non-Equilibrium Science (CNES) and the Thomas Young Centre (TYC).

\appendix

\medskip
\medskip
\medskip
\medskip
\medskip
\medskip
\medskip
\medskip

\noindent{\bf \LARGE Appendices}

\section{Derivation of the main result \eqref{scgfres} with \eqref{flowbeta}}\label{appder}

The derivation is based mainly on the assumptions of strong enough clustering of correlation functions of local observables, both in space and in time, along with a standard result from hydrodynamics, equation \eqref{resS}, which can be seen as a weak version of the hydrodynamic projection principle. Clustering is to be strong enough. For instance exponential clustering at large spacial separations can be shown rigorously in extremal KMS states associated with local hamiltonians, see e.g. \cite{Araki,Ruellebook,BratelliRobinson12}. Strong enough clustering in time is more difficult to prove, but expected to hold generically in many systems and states -- as explained in subsection \ref{ssectbreaking}, it is broken in some situations, leading to a breaking of the large deviation principle \eqref{ld}.

For completeness, we present various ways of proving the results, which involve different assumptions and principles. In one way of proving the emergence of a flow on the MES manifold, we make use of a theorem from \cite{Doyon2017} which shows that the pseudolocal charges, a concept originally introduced in \cite{ProsenPseudo1,ProsenPseudo2} (see \cite{IlievskietalQuasilocal}), form a Hilbert space that describe the tangent to the MES manifold as per \eqref{dbetaeq}. We require not only strong clustering in time, but also strong clustering along all rays $\xi=x/t$ in a neighbourhood of the $\xi=0$, expected to hold if no effective velocity lies within such a neighbourhood. In \cite{Doyon2017} the particular context of quantum statistical mechanics, specifically the quasi-local $C^*$ algebras, is taken, and a specific definition of MESs (which are in \cite{Doyon2017} referred to as GGEs) is used. This definition is in agreement with the one used here, in particular with \eqref{dbetaeq}, if we assume the tangent spaces, which may be different at different points along the flow, to have the same countable basis all along the flow, the $Q_i$'s. In generic, non-integrable system, we would in fact expect all tangent spaces to be finite dimensional, but this is difficult to prove.

In another way of proving the emergence of a flow on the MES manifold, we instead make use of a stronger version of the hydrodynamic projection principle, equation \eqref{proj}.  Hydrodynamic projection principles are nontrivial, but have been used successfully in the context of statistical fluid dynamics \cite{Spohn-book,SpohnNonlinear,1742-5468-2015-3-P03007,SciPostPhys.3.6.039}, see also \cite{doyoncorrelations} where an even stronger version of \eqref{proj} is explicitly used to derive correlation results verified numerically in \cite{10.21468/SciPostPhys.4.6.045}.

In addition, we make various technical assumptions, such as appropriate boundedness and differentiability assumptions for correlation functions. The precise specification of the assumptions of clustering, boundedness and differentiability is possible but would require a full mathematical framework -- such as that of $C^*$ algebras; we hope to come back to such matters in future works.

\subsection{Statement of the problem}\label{sappcumul}

Consider the expression \eqref{scgf} for the scaled cumulants $c_n$. Let us assume that correlation functions of the current $\losymbol{j}(0,t)$ cluster strongly enough at large time differences. Using stationarity of the state, standard arguments show that the scaled cumulants exist, and can be written in the form
\beq\label{cumul}
	c_n = \lim_{t_1\to\infty}\cdots
	\lim_{t_{n-1}\to\infty}\int_{-t_1}^{t_1}\dd s_1\cdots
	\int_{-t_{n-1}}^{t_{n-1}}\dd s_{n-1}
	\bra \losymbol{j}(0,s_{n-1}), \ldots, \losymbol{j}(0,s_1),\losymbol{j}(0,0)\ket_{\underline\beta}^{\rm c}
\eeq
where the limits are taken, in order, on $t_{n-1},\,t_{n-2},\,\ldots,t_{1}$, as written. Here the many-point connected, symmetrised correlation function on the right-hand side is a natural generalisation of  \eqref{connected}: we define $\bra\Or_1,\ldots,\Or_n\ket^{(\rm c)}_{\underline\beta}$ as the connected part of the expectation value of the normalised iterated anti-commutators
\beq\label{defconn}\begin{aligned}
	2^{-n+1}\{\Or_1, \{\ldots,
	\{\Or_{n-1},\Or_n\}\cdots\}\}
	\end{aligned}
\eeq
(where $\{\Or,\Or'\} = \Or\Or' + \Or'\Or$ is the anti-commutator).

Let us assume more generally that correlation functions involving the currents $\losymbol{j}(0,t)$ and local observables cluster strongly enough at large time differences. Then standard arguments show that (i) the state $\bra\cdots\ket^{(\lambda)}$ defined by the series expansion in $\lambda$ of \eqref{eq0} has a nonzero radius of convergence for any local observable $\Or(x,t)$ (the radius may depend on the observable), (ii) for any local observable $\Or(x,t)$, \eqref{derstate} holds within the convergence region, (iii) the series expansion in \eqref{scgf} for $F(\lambda)$ has a nonzero radius of convergence, and (iv) the result \eqref{Fder} holds within the convergence region.

There are then two ways -- which are equivalent under an appropriate assumption of clustering at large time differences -- to define precisely the function $F(\lambda)$, in order to prove \eqref{scgfres} with \eqref{flowbeta}:
\bi
\item[I.] We may define $F(\lambda)$ as the solution to \eqref{Fder} with $F(0)=0$, where the state $\bra\cdots\ket^{(\lambda)}$ is defined as the solution to \eqref{derstate} with $\bra\cdots\ket^{(0)} = \bra\cdots\ket_{\underline\beta}$.
\item[II.] We may take the explicit form of the cumulants \eqref{cumul}, along with the second equation in \eqref{scgf}, as our working expression.
\ei
Below we give simple proofs \eqref{scgfres} with \eqref{flowbeta} under both definitions. The arguments presented in subsection \ref{ssectskewing} follow the proof that is natural under definition I; this proof is conceptual, but neglects technical difficulties about the manifold structure of MES. The proof under definition II is more explicit.

\subsection{Hydrodynamic projections}

The main assumption for this section is the {\em hydrodynamic projection principle}. This principle has various implementations. We first use a very weak version of it, which expresses Fourier transforms of dynamical two-point functions of conserved densities in terms of the static correlation matrix $\mathsf C$ and the flux Jacobian $\mathsf A$, in the long-wavelength, long-time limit. This is in fact a standard result in linear fluctuating hydrodynamics.

Define
\beq
	S_{ij}(k,t) = \int_{-\infty}^\infty \dd x\,\re^{-\ri kx}
	\bra \losymbol{q}_i(x,t)\losymbol{q}_j(0,0)\ket^{\rm c}_{\underline\beta}
\eeq
as well as the ``Euler scaling limit" of long wavelengths and long times,
\beq\label{Sijeul}
	S_{ij}^{\rm Eul}(kt) = \lim_{k\to0,\,t\to\infty\atop kt\ \mbox{\tiny fixed}} S_{ij}(k,t).
\eeq
The weak version of the hydrodynamics projection principle that we assume is that the time dependence of $S_{ij}(k,t)$ takes the form
\beq\label{resS}
	S_{ij}^{\rm Eul}(kt) = \lt(\re^{-\ri kt \mathsf A}\mathsf C\rt)_{ij}.
\eeq

We need one additional assumption. As we will present two alternative arguments, this additional assumption takes two different forms, depending on the argument used.

In one form, we require that there exist a $k\in(0,1)$ such that clustering of correlation functions of local observables in time is strong enough along all rays $\xi=x/t$ in the interval $[-k,k]$,
\beq\label{condclus}
	\exists k\in(0,1)\ :\ \bra\Or(\xi t,t),\Or'(0,0)\ket^{\rm c}_{\underline\beta} \to 0\quad \mbox{fast enough as } t\to\infty\ \forall
	\ \xi\in[-k,k].
\eeq
We believe exponential clustering with uniform exponent would be sufficient, but it is not necessary.

In the other form, a stronger version of the hydrodynamic projection principle is required to hold: Fourier transforms of two-point functions involving a conserved density and an arbitrary local observable are expressed as Fourier transforms of conserved density two-point functions. Define, for local observables $\Or(x,t)$,
\beq
	S_{i\Or}(k,t) = \int_{-\infty}^\infty \dd x\,\re^{-\ri kx}
	\bra \losymbol{q}_i(x,t),\Or(0,0)\ket^{\rm c}_{\underline\beta},
\eeq
and the Euler scaling limit $S_{i\Or}^{\rm Eul}(kt)$ as in \eqref{Sijeul}. Then the stronger version is
\beq\label{proj}
	S_{i\Or}^{\rm Eul}(kt) = 
	\sum_{j,j'}S_{ij}^{\rm Eul}(kt)(\mathsf C^{-1})^{jj'} (\losymbol{q}_{j'},\Or)_{\underline\beta},
\eeq
where the inner product is defined in \eqref{innerproduct}. Equation \eqref{proj} represents the idea that correlations, at large scales, are produced by propagation of ballistically transported quantities, and thus two-point functions of the local observables are fully determined by evaluating their overlap with the conserved charges and propagating conserved densities. The static correlation matrix provides the metric in the space of conserved densities to use for the completeness relation. Below we provide a proof of the weaker version \eqref{resS} as a consequence of assumption \eqref{proj}, for completeness. We note that assumption \eqref{proj}, with \eqref{resS}, immediately implies
\beq\label{resS2}
	S_{i\Or}^{\rm Eul}(kt) = \lt(\re^{-\ri kt\mathsf A} (\underline{\losymbol{q}},\Or)_{\underline\beta}\rt)_i.
\eeq

With these assumptions, the main result of this subsection is to show that
\beq\label{resj}
	\int_{-\infty}^\infty \dd t\,
	\bra \losymbol{j}_i(0,t), \Or(0,0)\ket^{\rm c}_{\underline\beta}
	= \big(\sgn(\mathsf A)\,(\underline{\losymbol{q}},\Or)_{\underline\beta}\big)_i
	= -\sum_j \sgn(\mathsf A)_{i}^{~j}\frc{\p}{\p\beta_j}\bra\Or(0,0)\ket_{\underline\beta}.
\eeq
This implies in particular \eqref{qjA}, but is more general.

\medskip
\noindent {\em Proof of \eqref{resj}.} Note that \eqref{resS2}, specialised to $\Or(x,t) = \losymbol{q}_j(x,t)$, is the expression \eqref{resS}, our main assumption (weak hydrodynamic projection principle). Below, we show \eqref{resj} from \eqref{resS2}. Under the strong hydrodynamic projection principle \eqref{proj}, this completes the proof. Under the weaker one \eqref{resS}, we need one additional step: we need to show that there exists a conserved density $\t{\losymbol{q}}_i(x,t) = \sum_j \mathsf V_{i}^{~j}\losymbol{q}_i(x,t)$ such that
\beq\label{resjweak}
	\int_{-\infty}^\infty \dd t\,
	\bra \losymbol{j}_i(0,t), \Or(0,0)\ket^{\rm c}_{\underline\beta}
	= (\t{\losymbol{q}}_i,\Or)_{\underline\beta}
\eeq
for all local observables $\Or(x,t)$. Once this is shown, specialising it to $\Or(x,t)=\losymbol{q}_j(x,t)$ and using \eqref{resj} for $\Or(x,t) = \losymbol{q}_j(x,t)$, we determine that $\mathsf V_{i}^{~j} = \sgn(\mathsf A)_{i}^{~j}$, which completes the proof.

We first prove \eqref{resjweak}. The main idea is to show that $\int_{-\infty}^\infty \dd t\,\losymbol{j}_i(0,t)$ is a conserved pseudolocal charge. This is useful, as \cite[Defs 5.4, 6.2]{Doyon2017} says that the tangent space to a MES (as per \eqref{dbetaeq}) is the conserved subspace of the Hilbert space completion of the inner product induced by \eqref{connected}, and \cite[Thm 5.7]{Doyon2017} shows that this Hilbert space completion is in bijection with the space of pseudolocal charges. The precise definition of pseudolocal charges is given in \cite[Def 5.1]{Doyon2017} (we use the ``two-sided pseudolocal charges", see \cite[Sect 5.2]{Doyon2017}). Making contact with this definition, we must consider $t\mapsto J^{(t)} = \int_{-t/2}^{t/2} \dd s\,\losymbol{j}_i(0,t)$ (with an unimportant shift with respect to \eqref{Qt}) to form an infinite sequence of operators with increasing $t$, supported on increasing intervals whose length is proportional to $t$. The definition \cite[Def 5.1]{Doyon2017} asks for the support to be strict, while, under time evolution, it is known by the Lieb-Robinson theorem \cite{LiebRobinson,BHC06} that the support is only exponentially accurate. But by exponential accuracy, it is possible, using the techniques of \cite{BHC06}, to approximate the time-evolved fields by observables with strict supports, and modify the sequence in order for the strict support to grow linearly and converge to the same object. Here for simplicity we assume this has been done and that $J^{(t)}$ has strict support growing linearly with $t$. We must then check three requirements for the large-$t$ limit of $J^{(t)}$ to form, in the sense of \cite[Def 5.1]{Doyon2017}, a pseudolocal charge: (I) that $\bra J^{(t)},J^{(t)}\ket^{\rm c}_{\underline\beta}$ has a growth that is bounded linearly in $t$; (II) that $\lim_{t\to\infty}\bra J^{(t)},\Or(x,0)\ket^{\rm c}$ exists for all local observables $\Or(x,0)$; and (III) that the result of the latter is independent of $x$. For the latter point, in fact, a strong enough independence must hold not just in the limit: that there exists $k\in(0,1)$ such that the supremum of the difference $|\bra J^{(t)},\Or(x,0)\ket^{\rm c}-\bra J^{(t)},\Or(y,0)\ket^{\rm c}|$ within the region $x,y\in[-kt,kt]$ tend to zero as $t\to\infty$. As in the large-$t$ limit it is clear that $J^{(t)}$ is conserved, with these three requirements, \eqref{resjweak} follows from \cite[Thm 5.7]{Doyon2017}.

The first two requirements are immediate from strong enough clustering in time. The last one follows from the conservation laws \eqref{conslaw}, and strong enough clustering along all rays $\xi=x/t$ with $|\xi|\in[0,k]$, as expressed in \eqref{condclus}. Indeed, the conservation laws allow us bound the supremum of the difference by
\beq
	\int_{-kt}^{kt} \dd x\,|\bra\losymbol{q}_i(x,t),\Or(0,0)\ket^{\rm c}_{\underline\beta}| =
	t\int_{-k}^{k} \dd \xi\,|\bra\losymbol{q}_i(t\xi,t),\Or(0,0)\ket^{\rm c}_{\underline\beta}|.
\eeq
If clustering is strong enough along all rays $\xi=x/t\in[-k,k]$ (and using the dominated convergence theorem), the large-$t$ limit vanishes. This completes the proof of \eqref{resjweak}.

We now prove \eqref{resj} from \eqref{resS2}.
Let us take some $t>0$, and consider a function $f(x)$ whose derivative is $f'(x) \propto e^{-\mu x^2}$ for some $\mu>0$, with the normalisation condition
\beq\label{norm}
	\int_{-\infty}^\infty \dd x\,f'(x)=1,\qquad
	f(\infty)=-f(-\infty) =\frc12.
\eeq
Then using \eqref{resS2} as well as the dominated convergence theorem, we have
\beq\label{Seq}
	\lim_{\lambda\to\infty} \int_{-\infty}^\infty \dd x\,f(x/\lambda)
	\bra \losymbol{q}_i(x,\lambda t)\Or(0,0)\ket^{\rm c}_{\underline\beta} = \lt(f(t\mathsf A)(\underline{\losymbol{q}},\Or)_{\underline\beta}\rt)_{i}.
\eeq
The equation also holds with $t\mapsto-t$. By the conservation laws,
\beq
	\losymbol{q}_i(x,t) -
	\losymbol{q}_i(x,-t) = -\p_x
	\int_{-t}^t \dd s\,\losymbol{j}_i(x,s).
\eeq
By clustering in space, we can use integration by parts, and we find
\beq\begin{aligned}
	&\lim_{\lambda\to\infty} \int_{-\infty}^\infty \dd x\,f(x/\lambda)
	\bra \big(\losymbol{q}_i(x,\lambda t)-\losymbol{q}_i(x,-\lambda t)\big)\Or(0,0)\ket^{\rm c}_{\underline\beta} \\
	&\qquad\qquad=
	\lim_{\lambda\to\infty} \int_{-\infty}^\infty \dd x\,f'(x) \int_{-\lambda t}^{\lambda t}\dd s\,
	\bra \losymbol{j}_i(\lambda x,s)\Or(0,0)\ket^{\rm c}_{\underline\beta}.
	\end{aligned}\label{int}
\eeq
Again by the conservation laws,
\beq
	\int_{-t}^t \dd s\,\losymbol{j}_i(x,s)=
	\int_{-t}^t \dd s\,\losymbol{j}_i(0,s)
	- \int_0^x \dd y\,\lt(\losymbol{q}_i(y,t)
	-\losymbol{q}_i(y,-t)\rt).
\eeq
Inserting this in the right-hand side of \eqref{int} and using \eqref{Seq}, we obtain
\beq\begin{aligned}
	\big(f(t\mathsf A)(\underline{\losymbol{q}},\Or)_{\underline\beta}-f(-t\mathsf A)(\underline{\losymbol{q}},\Or)_{\underline\beta}\big)_{i}
	& =
	\lim_{\lambda\to\infty} \Bigg[
	\int_{-\infty}^\infty \dd x\,f'(x)\int_{-\lambda t}^{\lambda t} \dd s\,
	\bra \losymbol{j}_i(0,s)\Or(0,0)\ket^{\rm c}_{\underline\beta} \ - \\&\quad
	\int_{-\infty}^\infty \dd x\,f'(x)
	\int_0^{\lambda x} \dd y\,
	\bra (\losymbol{q}_i(y,\lambda t)
	-\losymbol{q}_i(y,-\lambda t))\Or(0,0)\ket^{\rm c}_{\underline\beta}
	\Bigg].
	\end{aligned}
\eeq
Using the first equation in \eqref{norm} to evaluate the first term on the right-hand side, and simplifying the integral in the second term, we find
\beq\begin{aligned}
	& \int_{-\infty}^{\infty} \dd s\,
	\bra \losymbol{j}_i(0,s)\Or(0,0)\ket^{\rm c}_{\underline\beta} \\ &  =
	\big(f(t\mathsf A)(\underline{\losymbol{q}},\Or)_{\underline\beta}-f(-t\mathsf A)(\underline{\losymbol{q}},\Or)_{\underline\beta}\big)_{ij}
	+
	\lim_{\lambda\to\infty}
	\int_{-\infty}^\infty \dd y\,g(y/\lambda)
	\bra (\losymbol{q}_i(y,\lambda t)
	-\losymbol{q}_i(y,-\lambda t))\Or(0,0)\ket^{\rm c}_{\underline\beta}
	\end{aligned}\label{int4}
\eeq
where, using the second equation in \eqref{norm},
\beq\label{gy}
	g(y) = \int_{y}^{\sgn(y)\infty} \dd x\,f'(x) = \frc{\sgn(y)}2-f(y).
\eeq
The second term on the right-hand side of \eqref{int4} can be evaluated by \eqref{Seq}. Simplifying by using \eqref{gy} and recalling that $t>0$, we obtain \eqref{resj}. \eproof

\medskip
\noindent {\em Proof of \eqref{resS} from \eqref{proj}.} First note that by symmetry (in particular, recall definition \eqref{connected}), assumption \eqref{proj} implies
\beq\label{proj2}
	S_{\Or i}^{\rm Eul}(kt) = 
	\sum_{j,j'} (\Or,\losymbol{q}_j)_{\underline\beta} (\mathsf C^{-1})^{jj'}S_{j'i}^{\rm Eul}(kt)
\eeq
where $S_{\Or i}^{\rm Eul}(kt)$ is the Euler scaling limit of $S_{\Or i}(k,t) = \int_{-\infty}^\infty \dd x\,\re^{-\ri kx}\bra \Or(x,t)\losymbol{q}_i(0,0)\ket^{\rm c}_{\underline\beta}$. Using the conservation laws \eqref{conslaw} and integration by parts, we obtain
\beq
	\p_t S_{ij}(k,t) = -\ri k S_{\losymbol{j}_i\,j}(k,t).
\eeq
From \eqref{proj2}, we then find in the Euler scaling limit, assuming that we can exchange the derivative and the limit,
\beqa
	\p_t S_{ij}^{\rm Eul}(kt) &=& -\ri k
	\sum_{\ell,\ell'} (\losymbol{j}_i,\losymbol{q}_\ell)_{\underline\beta} (\mathsf C^{-1})^{\ell \ell'}S_{\ell'j}^{\rm Eul}(kt) \n
	&=& -\ri k
	\sum_{\ell,\ell'} \frc{\p\mathsf{j}_i}{\p\beta_\ell}\,\frc{\p\beta_\ell}{\p\mathsf{q}_{\ell'}}\,S_{\ell'j}^{\rm Eul}(kt) \n
	&=& -\ri k
	\sum_{\ell'} A_i^{~\ell'}\,S_{\ell'j}^{\rm Eul}(kt) 
\eeqa
whose solution, with the initial condition $S_{ij}^{\rm Eul}(0) = \mathsf C_{ij}$, is \eqref{resS}. \eproof

\subsection{$\lambda$-flow}

We define the ``Lie derivative" ${\cal L} \bra\cdots\ket_{\underline\beta}$, at the point $\bra\cdots\ket_{\underline\beta}$ in the MES manifold characterised by the Lagrange parameters $\underline\beta$, by
\beq\label{lieder}
	{\cal L} \bra\Or\ket_{\underline\beta} = \int_{-\infty}^\infty \dd t\,\bra \losymbol{j}(0,t),\Or\ket_{\underline\beta}^{\rm c}.
\eeq
Leibniz's rule fixes its form on products of expectations, and in particular on connected, symmetrised correlation functions, a combinatoric analysis from the definition given around \eqref{defconn} gives
\beq\label{liederc}
	{\cal L} \bra\Or_1,\ldots,\Or_n\ket^{(\rm c)}_{\underline\beta} = \int_{-\infty}^\infty \dd t\,\bra \losymbol{j}(0,t),\Or_1,\ldots,\Or_n\ket_{\underline\beta}^{\rm c}.
\eeq
As a consequence of \eqref{resj},
\beq\label{loa}
	{\cal L}  \bra\Or\ket_{\underline\beta} 
	= -\sum_i \sgn(\mathsf A)_{i_*}^{~i}\frc{\p}{\p\beta_i}\bra\Or\ket_{\underline\beta},
\eeq
and, by Leibniz's rule,
\beq\label{loac}
	{\cal L}\bra \Or_1,\ldots,\Or_n\ket^{\rm c}_{\underline\beta}
	= -\sum_i \sgn(\mathsf A)_{i_*}^{~i}\frc{\p}{\p\beta_i}
	\bra \Or_1,\ldots,\Or_n\ket^{\rm c}_{\underline\beta}.
\eeq

Let us show \eqref{scgfres} with \eqref{flowbeta} using definition I of subsection \ref{sappcumul}. First \eqref{loa} shows that the Lie derivative on the MES manifold lies within the tangent space of the manifold. Therefore, by \eqref{derstate}, the flow $\lambda\mapsto \bra\cdots\ket^{(\lambda)}$ is that along the direction set by the Lie derivative ${\cal L}\bra\cdots\ket_{\underline\beta}$, starting at some point $\underline\beta(0)$, which lies entirely on the MES manifold. Hence, it can be characterised by $\lambda\mapsto \underline\beta(\lambda)$, which solves \eqref{flowbeta}:
\beq\label{lder}
	\bra\Or(0,0)\ket^{(\lambda)} = \bra\Or(0,0)\ket_{\underline\beta(\lambda)}.
\eeq
Finally, the differential equation \eqref{Fder} with $F(0)=0$ shows \eqref{scgfres}. \eproof

On the other hand, let us show \eqref{scgfres} with \eqref{flowbeta} using definition II of subsection \ref{sappcumul}, from the explicit expressions of the cumulants $c_n$ given in \eqref{cumul}. It is sufficient to show that
\beq\label{toshow}
	c_n = \frc{\dd^{n-1}}{\dd\lambda^{n-1}} \bra \losymbol{j}(0,0)\ket_{\underline\beta(\lambda)}\Big|_{\lambda=0}
\eeq
where $\underline\beta(\lambda)$ solves \eqref{flowbeta}. We show this by induction. The induction is on the statement that
\beq\label{induc}\begin{aligned}
	&\frc{\dd^{n-1}}{\dd\lambda^{n-1}} \bra \losymbol{j}(0,0)\ket_{\underline\beta(\lambda)}  \\ 
	&= \lim_{t_1\to\infty}\cdots
	\lim_{t_{n-1}\to\infty}\int_{-t_1}^{t_1}\dd s_1\cdots
	\int_{-t_{n-1}}^{t_{n-1}}\dd s_{n-1}
	\bra \losymbol{j}(0,s_{n-1}), \ldots, \losymbol{j}(0,s_1),\losymbol{j}(0,0)\ket_{\underline\beta(\lambda)}^{\rm c}.
	\end{aligned}
\eeq
In particular, this implies \eqref{toshow}. If \eqref{induc} holds for $n=m$, then we have
\beq\begin{aligned}
	&\frc{\dd^{m}}{\dd\lambda^{m}} \bra \losymbol{j}(0,0)\ket_{\underline\beta(\lambda)} \\
	&=\frc{\dd}{\dd\lambda} 
	\lim_{t_1\to\infty}\cdots
	\lim_{t_{m-1}\to\infty}\int_{-t_1}^{t_1}\dd s_1\cdots
	\int_{-t_{m-1}}^{t_{m-1}}\dd s_{m-1}
	\bra \losymbol{j}(0,s_{m-1}), \ldots, \losymbol{j}(0,s_1),\losymbol{j}(0,0)\ket_{\underline\beta(\lambda)}^{\rm c}.
	\end{aligned}
\eeq
Assuming that we can exchange the limits and integrals with the $\lambda$-derivative, this gives
\beq\begin{aligned}
	&\frc{\dd^{m}}{\dd\lambda^{m}} \bra \losymbol{j}(0,0)\ket_{\underline\beta(\lambda)} \\
	&=
	\lim_{t_1\to\infty}\cdots
	\lim_{t_{m-1}\to\infty}\int_{-t_1}^{t_1}\dd s_1\cdots
	\int_{-t_{m-1}}^{t_{m-1}}\dd s_{m-1}
	\frc{\dd}{\dd\lambda} 
	\bra \losymbol{j}(0,s_{m-1}), \ldots, \losymbol{j}(0,s_1),\losymbol{j}(0,0)\ket_{\underline\beta(\lambda)}^{\rm c}\\
	&=
	\lim_{t_1\to\infty}\cdots
	\lim_{t_{m-1}\to\infty}\int_{-t_1}^{t_1}\dd s_1\cdots
	\int_{-t_{m-1}}^{t_{m-1}}\dd s_{m-1}
	(-)\times \\ &\qquad\qquad\qquad\qquad \times\,\sum_i \sgn(\mathsf A)_{i_*}^{~i}\frc{\p}{\p\beta_i}
	\bra \losymbol{j}(0,s_{m-1}), \ldots, \losymbol{j}(0,s_1),\losymbol{j}(0,0)\ket_{\underline\beta(\lambda)}^{\rm c}\\
	&=
	\lim_{t_1\to\infty}\cdots
	\lim_{t_{m-1}\to\infty}\int_{-t_1}^{t_1}\dd s_1\cdots
	\int_{-t_{m-1}}^{t_{m-1}}\dd s_{m-1}
	{\cal L}
	\bra \losymbol{j}(0,s_{m-1}), \ldots, \losymbol{j}(0,s_1),\losymbol{j}(0,0)\ket_{\underline\beta(\lambda)}^{\rm c}\\
	&=
	\lim_{t_1\to\infty}\cdots
	\lim_{t_{m}\to\infty}\int_{-t_1}^{t_1}\dd s_1\cdots
	\int_{-t_m}^{t_{m}}\dd s_{m}
	\bra \losymbol{j}(0,s_{m}), \ldots, \losymbol{j}(0,s_1),\losymbol{j}(0,0)\ket_{\underline\beta(\lambda)}^{\rm c}
	\end{aligned}
\eeq
where on the third line we used \eqref{flowbeta} (along with Leibniz's rule), on the fourth \eqref{loac}, and on the fifth \eqref{liederc}. This show \eqref{induc} for $n=m+1$. Since \eqref{induc} holds by definition for $n=1$, this completes the proof. \eproof

\subsection{Second cumulant}\label{sappc2}

The second cumulant was shown in \cite{1742-5468-2015-3-P03007,SciPostPhys.3.6.039} to satisfy the sum rule
\beq
	c_2 = \lim_{t\to\infty} \int \dd x\,|x| S_{i_*i_*}(x,t).
\eeq
Using the Euler-scale expression \eqref{resS}, this gives
\beq\label{c2sumrule}
	c_2 = (|\mathsf A|\mathsf C)_{i_*i_*}
\eeq
where $|A|=\sgn(A)A$. On the other hand, from our result \eqref{scgfres} with \eqref{flowbeta}, we have
\beq
	c_2 = \frc{\dd \mathsf{j}_{i_*}}{\dd\lambda}
	= \sum_{i,j} \frc{\p\beta^i}{\p\lambda} \frc{\p\mathsf{q}_j}{\p\beta^i} \frc{\p\mathsf{j}_{i_*}}{\p\mathsf{q}_j}
	= \sum_{ij}\sgn(\mathsf A)_{i_*}^{~i}\mathsf C_{ij}\mathsf A_{i_*}^{~j} 
	= (\sgn(\mathsf A)\mathsf A\mathsf C)_{i_*i_*}
\eeq
where in the last step we used \eqref{AC}. This agrees with \eqref{c2sumrule}.

\section{Euler hydrodynamics}\label{appeuler}

\subsection{Standard results}

Consider a non-stationary, inhomogeneous state $\bra\cdots\ket$ of the system. In Euler hydrodynamics, one assumes the every local average at space-time point $x,t$ can be approximated by a local MES, that depends on $x,t$ but not on the observable whose average is taken,
\beq\label{hydroapprox}
	\bra \Or(x,t)\ket \approx \bra\Or(0,0)\ket_{\underline\beta(x,t)}
\eeq
(since the MES is homogeneous and stationary, one can put the observable at $0,0$ on the right-hand side). Physically, there are fluid cells, which are large compared to microscopic scales but small compared to variation scales of the states, in which the state has, to a good approximation, maximised entropy, and is very nearly homogeneous and stationary. Writing the conservation laws \eqref{conslaw} in average form within the state $\bra\cdots\ket$ and using the approximation \eqref{hydroapprox}, one obtains\footnote{A more precise derivation can be obtained by re-writing the conservation laws in integral form, and assuming that, in the appropriate Euler scaling limit, \eqref{hydroapprox} holds uniformly enough.}
\beq
	\p_t \mathsf{q}_i(x,t) + \p_x \mathsf{j}_i(x,t) = 0
\eeq
where $\mathsf{q}_i(x,t) = \bra \losymbol{q}_i(0,0)\ket_{\underline\beta(x,t)}$ and $\mathsf{j}_i(x,t) = \bra \losymbol{q}_i(0,0)\ket_{\underline\beta(x,t)}$. This can be re-written using the flux Jacobian \eqref{Amatrix},
\beq\label{eulerquasilinear}
	\p_t \mathsf{q}_i(x,t) + \sum_{j} \mathsf A_{i}^{~j}(x,t)\p_x \mathsf{q}_j(x,t) = 0.
\eeq
Since $\mathsf{q}_i(x,t)$ form a system of coordinates for the MES at $x,t$, these are equations of motion of the space-time dependent MES -- they are the Euler hydrodynamic equations corresponding to the dynamical system of interest.

It is known \cite{Spohn-book,SpohnNonlinear,1742-5468-2015-3-P03007,SciPostPhys.3.6.039} (see also the explicit proof in \cite{PhysRevX.6.041065}) that there exists a generating function $g$ for the currents,
\beq\label{jgapp}
	\mathsf{j}_i = -\frc{\p g}{\p\beta^i}.
\eeq
The function $g$ is a generating function for the average currents in MES, much like the specific free energy is for the average densities. Equation \eqref{jgapp} is a consequence of the fact that $-\p \mathsf{j}_i/\p\beta^j  = \int \dd x\,\bra \losymbol{q}_j(x,0),\losymbol{j}_i(0,0)\ket_{\underline\beta}^{\rm c}$ is symmetric under $i\leftrightarrow j$, which can be shown by using the conservation laws and integration by parts. Note that changing variables to $\underline{\mathsf{q}}$, this symmetry also immediately implies the important relation
\beq\label{AC}
	\mathsf A\mathsf C = \mathsf C\mathsf A^T \quad\mbox{or} \quad \sum_j\mathsf A_i^{~j}\mathsf C_{jk} = \sum_j\mathsf C_{ij}\mathsf A_k^{~j}
\eeq
involving the static correlation matrix $\mathsf C = \frc{\p\underline{\mathsf{q}}}{\p\underline\beta}$ (see \eqref{Cmatrix}).

The normal coordinates of Euler hydrodynamics are a different system of coordinates, $\underline{\mathsf{q}}\mapsto \underline n$, which diagonalise the flux Jacobian. That is, there is a diagonal matrix $v^{\rm eff}$, with diagonal elements $v^{\rm eff}_i$, such that
\beq\label{veffdiagq}
	v^{\rm eff} = \frc{\p\underline{n}}{\p\underline{\mathsf{q}}} \mathsf A \frc{\p\underline{\mathsf{q}}}{\p\underline{n}}\quad\mbox{or}\quad
	 \delta_{i,j}v^{\rm eff}_i = \sum_{k,l} \frc{\p n_i}{\p \mathsf{q}_k}
	\mathsf A_{k}^{~l}\frc{\p \mathsf{q}_l}{\p n_j}.
\eeq
The quantities $v_i^{\rm eff}$ are the ``effective velocities" of the normal modes in the fluid, nonlinear versions of the sound velocity. Changing coordinates, one then obtains
\beq
	\p_t n_i(x,t) + v^{\rm eff}_i(x,t) \p_x n_i(x,t) = 0.
\eeq

It is also possible to re-write all equations in terms of the coordinates given by the Lagrange parameters $\underline\beta$. Using \eqref{eulerquasilinear} and \eqref{AC}, we find
\beq\label{eulerquasilinearbeta}
	\p_t \beta^i(x,t) + \sum_{j} \p_x \beta^j(x,t)\mathsf A_{j}^{~i}(x,t) = 0.
\eeq
Further, we also have
\beq
	v^{\rm eff} = \frc{\p\underline{n}}{\p\underline\beta}
	\frc{{\p\underline\beta}}{\p\underline{\mathsf{q}}} \mathsf A \frc{\p\underline{\mathsf{q}}}{\p\underline\beta} \frc{\p\underline\beta}{\p\underline{n}}
	=\frc{\p\underline{n}}{\p\underline\beta}
	\mathsf C^{-1}\mathsf  A \mathsf C\frc{\p\underline\beta}{\p\underline{n}}.
\eeq
and using \eqref{AC} we get
\beq\label{veffdiag}
	v^{\rm eff} = \frc{\p\underline{n}}{\p\underline\beta}
	\mathsf A^T\frc{\p\underline\beta}{\p\underline{n}}.
\eeq

\subsection{The Riemann problem in linear hydrodynamics}\label{sappfree}

Consider a linear Euler hydrodynamics, with $\mathsf A_i^{~ j}$ independent of the state. We can use \eqref{eulerquasilinearbeta} with flux Jacobian independent of $x,t$: 
\beq
	\p_t \beta^i(x,t) + \sum_{j} \p_x \beta^j(x,t)\mathsf A_{j}^{~i} = 0.
\eeq
Let us assume that $\mathsf A_j^{~i}$ is diagonalisable by a similarity transformation. Consider its right-eigenvectors:
\beq\label{Aeigen}
	\sum_i \mathsf A_j^{~ i}w_{ik} = v^{\rm eff}_kw_{jk}.
\eeq
Then we find normal modes by the linear coordinate transformation
\beq\label{nmode}
	n_k = \sum_i \beta^iw_{ik}
\eeq
and we have
\beq
	\p_t n_k(x,t) + v^{\rm eff}_k \p_x n_k(x,t) = 0.
\eeq

The Riemann problem is that of solving the Euler hydrodynamic equations with initial conditions giving by two distinct homogeneous states, one on the left, one on the right. We set
\beq
	\beta^i(x,0) = \lt\{\ba{ll} \beta^i_\rl & (x<0) \\
	\beta^i_\rr & (x>0).
	\ea\rt.
\eeq
Since both the initial condition in the Riemann problem and the Euler equation are invariant under simultaneous scaling $(x,t)\mapsto (\mu x\,\mu t)$, we may assume the solution to the Riemann problem to have this symmetry as well: all functions of space time are functions of $\xi = x/ t$ only. We therefore obtain
\beq
	(\xi-v^{\rm eff}_k)\p_\xi n_k(\xi;\underline\beta_\rl,\underline\beta_\rr) = 0.
\eeq
The solution is\footnote{This solution is smooth except for contact discontinuities. It does not contain shocks, hence does not  absorb (or generate) entropy, and is therefore expected to be the physically relevant solution.}
\beq\label{solfreemode}
	n_k(\xi;\underline\beta_\rl,\underline\beta_\rr) = \lt\{\ba{ll}
	n_{k;\rl} & (\xi<v^{\rm eff}_k) \\
	n_{k;\rr} & (\xi > v^{\rm eff}_k)
	\ea\rt.
\eeq
where $n_{k;\rl}$ and $n_{k;\rr}$ are the normal modes in the states specified by the Lagrange parameters $\underline\beta_\rl$ and $\underline\beta_\rr$, respectively. Let us consider the ray $\xi=0$, which is the relevant one for the transport statistics problem, and denote $n_k(0;\underline\beta_\rl,\underline\beta_\rr) = n_k(\underline\beta_\rl,\underline\beta_\rr)$. Let us consider the flow \eqref{flowbeta} for generating the transport statistics in the state specified by $n_k(\underline\beta_\rl,\underline\beta_\rr)$, and denote by $n_k(\underline\beta_\rl,\underline\beta_\rr;\lambda)$ the normal modes along the flow. Let us finally consider the solution \eqref{solfree} to the flow problem in the free case. Using \eqref{nmode} and \eqref{Aeigen}, this is
\beq
	n_k(\underline\beta_\rl,\underline\beta_\rr;\lambda) = n_k(\underline\beta_\rl,\underline\beta_\rr) - \lambda \sgn(v^{\rm eff}_k) w_{i_*k}.
\eeq
Using \eqref{solfreemode} we obtain
\beq
	n_k(\underline\beta_\rl,\underline\beta_\rr;\lambda) =
	n_k(\beta_\rl^\bullet - \lambda\delta_{i_*}^\bullet,\,\beta_\rr^\bullet + \lambda\delta_{i_*}^\bullet)
\eeq
which shows the general expression of the extended fluctuation relations \eqref{efrgeneral}.

\subsection{Some aspects of conformal hydrodynamics}

It is a simple matter to solve the diagonalisation problem for the flux Jacobian \eqref{dCFT A} of conformal hydrodynamics in arbitrary dimension. We find that the following combinations of the rest-frame temperature $T_\rre$ and the boost $\theta$ form normal modes:
\al{\label{hdCFTnormal}
n_+ =   T_\rre \re^{\frac{\theta}{\sqrt{d}}} ,\qquad
n_- =   T_\rre \re^{-\frac{\theta}{\sqrt{d}}}
}
with effective velocities $v^{\rm eff}_\pm$, respectively, as given in \eqref{xipm}.

It is also a simple matter to find the current generating functions \eqref{jgapp}. One can check that the currents $\mathsf{j}_1$ and $\mathsf{j}_2$ as given in \eqref{j1} and \eqref{j2} are generated as per \eqref{jgapp} by the function
\beq\label{hdCFTg}
	g = -T_\rre^d\sinh\theta.
\eeq

\section{Multi-parameter SCGF and normal mode decompositions} \label{appmany}

In certain cases, even beyond free models, we can obtain an explicit expression for $F(\lambda)$, where the integral in \eqref{scgfres} is performed in terms of the {\em normal modes} of the Euler hydrodynamics of the model. This result holds in hydrodynamic theories where the generating function for the currents, Eq.~\eqref{jgapp}, has a property of separation into normal modes, Eq. \eqref{gG} below. We do not know yet the full range of theories with this property, but it includes generalised hydrodynamics \cite{PhysRevX.6.041065}.

Consider the generating function for the currents \eqref{jgapp}. Arguments (see below) suggest that this function may in some situations separate into a sum of functions of the normal coordinates:
\beq\label{gG}
	g = \sum_i G_i(n_i).
\eeq
This decomposition holds in generalised hydrodynamics \cite{PhysRevX.6.041065}, but it is clear, from \eqref{hdCFTnormal} and \eqref{hdCFTg}, that it holds in conformal hydrodynamics if and only if $d=1$ (the ``trivial", linear case). In the cases where it holds, the SCGF is given by
\beq\label{SCGF}
	F(\lambda) = \sum_i \Big(\sgn\big(v^{\rm eff}_i(\lambda)\big)G_i(n_i(\lambda))\big) - 
	\sum_{a\in\{\pm\}}\sum_{\t\lambda\in \lambda_\star^{(i,a)}\cap I_\lambda}
	a\,G_i(n_i(\t\lambda))
	\Big)
\eeq
where $I_\lambda = [0,\lambda)$ if $\lambda>0$ and $(\lambda,0]$ if $\lambda<0$, and the sets $\lambda_\star^{(i,\pm)}$ are the turning points of the sign of the effective velocity,
\beq
	\lambda_\star^{(i,\pm)} = \{\t\lambda:v^{\rm eff}_i(\t\lambda) = 0,\,\t\lambda\p_{\t\lambda} v^{\rm eff}_i(\t\lambda) \gtrless 0\}.
\eeq
This parallels what is found in generalised hydrodynamics \cite{Myers-Bhaseen-Harris-Doyon}.

In order to show \eqref{SCGF}, we calculate, using \eqref{gG} and \eqref{jgapp},
\beqa
	\p_\lambda G_i(n_i(\lambda))
	&=& -\sum_{j,k}\frc{\p n_i}{\p \beta^k}\frc{\p \beta^k}{\p \lambda} 
	\frc{\p\beta^j}{\p n_i}
	\mathsf{j}_j \n
	&=& \sum_{j,k}\frc{\p n_i}{\p \beta^k}\sgn(\mathsf A^T)^k_{~i_*}
	\frc{\p\beta_j}{\p n_i}
	\mathsf{j}_j\n
	&=& \sgn(v^{\rm eff}_{i}) \frc{\p n_i}{\p \beta^{i_*}} \sum_{j}
	\frc{\p\beta^j}{\p n_i}
	\mathsf{j}_j
\eeqa
where in passing from the second to the third line we used \eqref{veffdiag}. Therefore, assuming we are away from the turning points of $v^{\rm eff}_i(\lambda)$, we have
\beq
	\p_\lambda F(\lambda)  = \sum_i \sgn(v^{\rm eff}_i) \p_\lambda G_i(n_i(\lambda))
	= \sum_{i,j}\frc{\p n_i}{\p \beta^{i_*}} 
	\frc{\p\beta^j}{\p n_i}
	\mathsf{j}_j = \mathsf{j}_{i_*}
\eeq
in agreement with \eqref{scgfres}. At the turning points of $v^{\rm eff}_i$, there are additional delta-function terms. One can verify that the second term in \eqref{SCGF} exactly cancels these terms.

In fact, using \eqref{dft} below, we can also write the multi-parameter SCGF $F(\underline\lambda)$ (see remark \ref{remaext}) as a function of the normal coordinates $\underline n = \underline n(\lambda)$ in a similar fashion, as
\beq
	F = \sum_i \Big(\sgn(v^{\rm eff}_i) G_i(n_i)
	- \sum_{\{a_j \in\{\pm\}\}_j}\sum_{\t {\underline n}\in n_\star^{(i,\underline a)}\atop  a_j (n_j-\t n_j)>0}
	G_i(\t n)
	\Big) + F_0
\eeq
where the sets $ n_\star^{(i,\underline a)}$ are the turning points of the sign of the effective velocity,
\beq
	n_\star^{(i,\underline a)} = \{\t {\underline n}:v^{\rm eff}_i(\t {\underline n}) = 0,\,a_j\p_{\t n_j} v^{\rm eff}_i(\t {\underline n}) \gtrless 0\}.
\eeq
The constant $F_0$ is such that at the original state $\underline n(\lambda=0)$, we recover $F(0)=0$.

We now provide an argument for the decomposition \eqref{gG}. Let us consider $g$ in \eqref{jgapp} as a function of the normal coordinates $\underline n$. We  argue, under certain assumptions (which may be  hard to verify), that $\p g/\p n_i$ is independent of $n_j$ for $j\neq i$. This would imply the decomposition \eqref{gG}.

For this argument, we consider the multi-parameter SCGF. Let us assume that there is a differentiable multi-parameter flow
\beq\label{mflow}
\frac{\p \beta^i }{\p \lambda_j} = -\sgn (\mathsf A)_{j}^{~i}
\eeq
and a differentiable SCGF $F(\underline\lambda)$ for the transport of all charges $Q_{j}$, each associated to $\lambda_j$, as per remark \ref{remaext}. We combine \eqref{scgfres} with \eqref{mflow} in this general situation, in order to obtain, assuming the matrix $\p\underline \beta/\p\underline\lambda$ to be invertible,
\beq\label{ert}
	\frc{\p F}{\p\underline\beta} \sgn(\mathsf A^T) = \frc{\p g}{\p{\underline\beta}}
\eeq
where $\frc{\p F}{\p\underline\beta}$ and $\frc{\p g}{\p{\underline\beta}}$ are to be seen as line vectors. Changing variables, we have
\beq
	\frc{\p F}{\p\underline n} \frc{\p\underline n}{\p\underline \beta}\sgn(\mathsf A^T) \frc{\p\underline \beta}{\p\underline n} = \frc{\p g}{\p{\underline n}}
\eeq
and using \eqref{veffdiag} this gives
\beq\label{dft}
	\frc{\p F}{\p\underline n}  = \frc{\p g}{\p{\underline n}} \sgn(v^{\rm eff}).
\eeq

We do not expect differentiability of $F$ at the points where $\sgn(v^{\rm eff})$ changes. However, away from these points, it is natural to assume that $F$ is differenitable. Consider therefore taking in \eqref{dft} another derivative with respect to the normal modes. Away from the points where the effective velocity changes sign, $\sgn(v^{\rm eff}_j)$ has zero derivative, and we obtain
\beq
	\frc{\p^2 F}{\p n_i \p n_j} = \frc{\p^2 g}{\p n_i \p n_j} \sgn(v^{\rm eff}_j).
\eeq
Since the left-hand side is symmetric (by differentiability of $F$), we find
\beq
	\frc{\p^2 g}{\p n_i \p n_j} (\sgn(v^{\rm eff}_i)-\sgn(v^{\rm eff}_j)) = 0.
\eeq
With $i\neq j$, in states where the signs of $v^{\rm eff}_i$ and $v^{\rm eff}_j$ are different, this implies
\beq\label{gnn}
	\frc{\p^2 g}{\p n_i \p n_j} = 0.
\eeq
Suppose that for each $i>j$, there exists a neighbourhood of states such that $\sgn(v^{\rm eff}_i)\neq \sgn(v^{\rm eff}_j)$. Then \eqref{gnn} will hold in all these neighbourhoods, for the corresponding $(i,j)$. Suppose also that for all $j$ and all $i$, the function $\p g/\p n_j$ is analytic in $n_i$ (in appropriate neighbourhoods of  $n_i$ such that $\underline n$ lies in the manifold of MES). Then, by analytic continuation, one would have \eqref{gnn} for all $\underline n$, and therefore \eqref{gG}.

One can verify that the multi-parameter flow \eqref{mflow} for the flux Jacobian of higher-dimensional CFT \eqref{dCFT A} is not consistent: it does not lead to differentiable Lagrange parameters as functions of the many parameters $\lambda_i$, at least in the region $|\theta|<\theta_{\rm s}$ (and we note that in the region $|\theta|>\theta_{\rm s}$, the effective velocities have the same sign). This is, technically, where the above argument fails in this case. It would be interesting to further study this situation.



\end{document}